\documentclass[12pt]{iopart}
\usepackage{graphicx}

\newcommand{\here}{\makebox(0,0)}
\newcommand{\bsigma}{\mbox{\boldmath $\sigma$}}

\newcommand{\btau}{\mbox{\boldmath $\tau$}}
\newcommand{\bxi}{\mbox{\boldmath $\xi$}}

\newcommand{\bm}{\mbox{\boldmath $m$}}

\newcommand{\bu}{\mbox{\boldmath $u$}}
\newcommand{\bv}{\mbox{\boldmath $v$}}

\newcommand{\bD}{\mbox{\boldmath $D$}}

\newcommand{\bT}{\mbox{\boldmath $T$}}
\newcommand{\bU}{\mbox{\boldmath $U$}}

\newcommand{\bS}{\mbox{\boldmath $S$}}

\newcommand{\pprime}{{\prime\prime}}
\newcommand{\R}{{\rm I\!R}}
\newcommand{\bigbra}{\left\langle}
\newcommand{\bigket}{\right\rangle}
\newcommand{\vsp}{\vspace*{3mm}}
\newcommand{\order}{ {\cal O}}

\newcommand{\arctanh}{ ~{\rm arctanh} }

\newcommand{\smallroom}{\rule[-0.1cm]{0cm}{0.3cm}}
\newcommand{\be}{\begin{equation}}
\newcommand{\ee}{\end{equation}}
\newcommand{\bd}{\begin{displaymath}}
\newcommand{\ed}{\end{displaymath}}
\newcommand{\bra}{\langle}
\newcommand{\ket}{\rangle}
\begin{document}

\title[{Diagonalization of replicated transfer matrices}]{Diagonalization of replicated transfer matrices for disordered Ising spin systems}
\author{T Nikoletopoulos and A C C Coolen}

\address{Department of Mathematics, King's College London, The Strand,
London WC2R 2LS, United Kingdom}

\begin{abstract}
We present an alternative procedure for solving the eigenvalue
problem of replicated transfer matrices describing disordered spin
systems with (random) 1D nearest neighbor bonds and/or random
fields, possibly in combination with  (random) long range bonds.
Our method is based on transforming the original eigenvalue
problem for a $2^n\times 2^n$ matrix (where $n\to 0$) into an
eigenvalue problem for integral operators. We first develop our
formalism for the Ising chain with random bonds and fields, where
we recover known results. We then apply our methods to models of
spins which interact simultaneously via a one-dimensional ring and
via more complex long-range connectivity structures, e.g.
$1+\infty$ dimensional neural networks and `small world' magnets.
Numerical simulations confirm our predictions satisfactorily.
\end{abstract}

\ead{theodore@mth.kcl.ac.uk,tcoolen@mth.kcl.ac.uk}

\section{Introduction}

The replica formalism, see e.g. \cite{spinglass},  has proven to
be a very powerful tool in the study of both statics and dynamics
of disordered systems. In statics the presence of frozen disorder
in the Hamiltonian makes a direct equilibrium statistical
mechanical analysis very difficult. Instead, one starts from the
key assumption (supported by numerical, experimental and sometimes
analytical evidence) that in the thermodynamic limit the free
energy per degree of freedom in such systems is self averaging,
i.e. identical to its disorder average for any given realization
of the disorder, with probability one. This property allows one to
focus on the evaluation of the disorder-averaged free energy per
degree of freedom, which for a disordered system of $N$
interacting discrete spins $\sigma_i$ ($i=1\ldots N$) in
equilibrium at inverse temperature $\beta=T^{-1}$ is calculated
using the following identity:
\be
\label{REPLICA}
  \bar{f}
       =-\lim_{N\to\infty}\frac{1}{\beta N}\lim_{n\to 0}
           \frac{1}{n}\log\overline{Z^{n}}~~~~~~~~Z=\sum_{\bsigma}e^{-\beta
           H(\bsigma)}
\ee Here $\bsigma=(\sigma_1,\ldots,\sigma_N)$,  $H(\bsigma)$ is
the Hamiltonian, and $\overline{\ldots\smallroom}$ denotes an
average over the disorder variables of the model under
consideration. The replica method involves changing the order of
the limits $n\to 0$ and $N\to\infty$, and subsequently writing the
$n$-th moment of the partition function $Z$ in terms of $n$ copies
(or replicas) of the original system. The disorder average then
converts the original problem of $n$ independent but disordered
systems into a new problem for $n$ coupled but disorder-free ones.
In the limit $N\to\infty$ this new non-disordered problem can
often be solved with conventional methods, e.g saddle point
integration. The limit $n\to 0$ has to be taken in the final
result. This procedure has over the years been applied with great
success to many families of mostly range-free or mean-field
models.

Application of the replica formalism to finite dimensional spin
models with disordered bonds and/or fields leads to the notion of
replicated transfer matrices \cite{weigt-monasson96,weigt98}. For
disordered one dimensional Ising spin chains, for instance, the
replica method effectively replaces an expression for  the free
energy in terms of products of $2\times 2$ random transfer
matrices (see e.g. \cite{Yeomans} for transfer matrix methods) by
an expression for an $n$-replicated chain without disorder but
with a more complicated $2^n\times 2^n$ transfer matrix which
couples the $n$ replicas at each site of the chain. In the
thermodynamic limit one first has to find the largest eigenvalue
of this replicated transfer matrix, and subsequently find its
analytic continuation for $n\to 0$. It was shown in
\cite{weigt-monasson96} that for the one dimensional Ising model
with random bonds and fields this procedure yields the results
obtained earlier by other techniques, see e.g.
\cite{BrandtGross78,BruinsmaAeppli83,derrida-hilhorst83}.
Moreover, it was found that
the smaller eigenvalues of the replicated transfer matrix contain
information about disorder-averaged two-spin connected correlation
functions.

In this paper we show how the replicated transfer matrix of
disordered Ising models can be diagonalized, by using a particular
form for the eigenvectors which transforms the original eigenvalue
problem into an eigenvalue problem  for integral operators. We
believe our method to have a number of possible advantages. It
appears more direct and explicit than existing approaches, it can
be generalized in a straightforward manner to situations with RSB
(which could for instance be induced by super-imposed long-range
bonds), and it does not rely on the limit $n\to 0$ being taken (so
that it can also be used for finite $n$ replica calculations
describing models where the disorder is not truly frozen but
evolving on very large time scales, in the sense of
\cite{PCS1,DotsFranzMezard,JABCP,CoolenUezu}).

 We first apply our
ideas to Ising chains with random bonds and fields, where we can
compare the results obtained with our method to those obtained
earlier by others. Furthermore, mathematically one may express
replicated transfer matrices of models which are not purely
one-dimensional (due to super-imposed long range bonds) in terms
of those corresponding to random field chains, with the statistics
of the random fields mediating the mean-field effect of the long
range bonds on a given site. We apply our equations to two
examples of such models with one-dimensional and long-range bonds:
the $1+\infty$ attractor neural networks of
\cite{skantzos-coolen00,skantzos-coolen01}, and the `small-world'
ferromagnet of \cite{nik-coolen04}, and show how one can use our
methods  to calculate various thermodynamic quantities.

\section{Definitions}

In this paper we will deal with disordered Ising spin systems in
thermal equilibrium, of size $N$ and with microscopic states
written as $\bsigma=(\sigma_1,\ldots,\sigma_N)\in\{-1,1\}^N$. More
specifically, we will analyze the following three models, by
diagonalizing the replicated transfer matrices which they
generate: the disordered Ising chain (DIC) as in
\cite{BrandtGross78,BruinsmaAeppli83}, the
$(1+\infty)-$dimensional attractor neural network (ANN) as in
\cite{skantzos-coolen00,skantzos-coolen01} and the `small world'
ferro-magnet (SWM) of \cite{nik-coolen04}, which are defined by
the Hamiltonians
\begin{eqnarray}
  H_{\rm DIC}(\bsigma)&=&-\sum_{i}J_{i}\sigma_{i}\sigma_{i+1}-\sum_{i}\theta_{i}\sigma_{i}
\label{eq:1DH}
\\
  H_{\rm ANN}(\bsigma)&=&
  -J_{s}\sum_{i}\sigma_{i}\sigma_{i+1}(\bxi_{i}\cdot\bxi_{i+1})
   -\frac{J_{\ell}}{N}\sum_{i<j}\sigma_{i}\sigma_{j}(\bxi_{i}\cdot\bxi_{j})
   \label{eq:NNH}
\\
  H_{\rm SWM}(\bsigma)&=& -J_{0}\sum_{i}\sigma_{i}\sigma_{i+1}-\frac{J}{c}
   \sum_{i<j}c_{ij}\sigma_{i}\sigma_{j}
   \label{eq:SWH}
\end{eqnarray}
 In (\ref{eq:1DH}) we have a 1D spin
chain with independently identically distributed random bonds and
fields $\{J_i,\theta_i\}$ at each site, drawn from some joint
distribution $p(J,\theta)$. We will abbreviate
$\int\!\mathrm{d}J\mathrm{d}\theta~p(J,\theta)f(\theta,J) =\bra
f(J,\theta)\ket_{J,\theta}$. In (\ref{eq:NNH}) we have both 1D and
long-range random bonds, but their values are not independent. The
short- and long-range bonds take the values
$J_s(\bxi_i\cdot\bxi_{i+1})$ and $J_\ell N^{-1}(\bxi_i\cdot\bxi_j)$,
respectively, where the binary vectors
$\bxi_i=(\xi_i^1,\ldots,\xi_i^p)$ represent stored data and are
drawn randomly and independently from $\{-1,1\}^p$ (with uniform
probabilities). Finally, in (\ref{eq:SWH}) we have uniform 1D
ferromagnetic bonds of strength $J_0$, and  the randomness is
solely in the realization of the long range bonds. The latter are
also ferromagnetic, of strength $J/c$ if present, but constitute a
finitely connected Poissonian graph defined by dilution variables
$c_{ij}$ which for each pair $(i,j)$  are drawn independently from
$p(c_{ij})=\frac{c}{N}\delta_{c_{ij},1}+(1-\frac{c}{N})\delta_{c_{ij},0}$.
The average connectivity $c$ will remain finite in the
thermodynamic limit. In all three cases
(\ref{eq:1DH},\ref{eq:NNH},\ref{eq:SWH}) the 1D short-range
interactions are defined periodically.

At this stage, let us briefly recall from
\cite{weigt-monasson96,weigt98} how a replicated transfer matrix
emerges for the disordered Ising chain (\ref{eq:1DH}) upon
applying the replica identity (\ref{REPLICA}). Here one finds,
with $\alpha=1\ldots n$ and with the short-hand
$\bsigma_i=(\sigma_{i}^{1},\ldots,\sigma_{i}^{n})\in\{-1,1\}^n$,
\begin{eqnarray}
  \overline{Z^{n}}&=&
\sum_{\bsigma_1\ldots \bsigma_N}\prod_{i} \bigbra e^{\beta
J\sum_{\alpha}\sigma_{i+1}^{\alpha}\sigma_{i}^{\alpha}+\beta\theta\sum_{\alpha}
\sigma_{i}^{\alpha}}\bigket_{J,\theta} =\tr(\bT_{n}^{N})
\end{eqnarray}
with a $2^{n}\times 2^{n}$ matrix $\bT_{n}$ whose entries are
given by
\be
  T_{n}(\bsigma,\bsigma^\prime)=\bigbra
 e^{\beta J\sum_{\alpha}\sigma_{\alpha}\sigma_{\alpha}^\prime +\beta\theta\sum_{\alpha}\sigma_{\alpha}}\bigket_{J,\theta}
 \label{rtm_1d}
\ee
 One can thus find the disorder-averaged free energy per spin
in the usual manner, via (\ref{REPLICA}), by determining the
largest eigenvalue of the replicated transfer matrix $\bT_{n}$ for
integer $n$. The difficulty lies in the requirement to find an
analytic expression for this eigenvalue for {\em arbitrary}
integer $n$ (in contrast to non-disordered chains, where the
dimension is fixed from the start and usually small, and where
direct methods can therefore be employed such as calculating the
characteristic polynomial of the matrix and finding its zeros).

We will first develop our diagonalization method for the simplest
case, viz. the chain (\ref{eq:1DH}), and subsequently show that it
can also serve to generate the  solution of the other two models
(\ref{eq:NNH},\ref{eq:SWH}), which involve both short- and long
range bonds, by writing the transfer matrices of the latter two
models again in the form  (\ref{rtm_1d}), but with suitably
defined distributions $p(J,\theta)$ of local bonds and fields.

\section{Construction and properties of eigenvectors}
\label{sec:theory}

\subsection{A detour: the Ising chain without disorder}

 Let us first turn to the simplest
possible case: the 1D Ising chain with bonds $J$ and uniform
fields $\theta$ (without disorder), where we just have the
familiar transfer matrix
\be
\label{eq: 2by2matrix}
  T(\sigma,\sigma^\prime;\theta,J)=e^{\beta J\sigma\sigma^\prime+\beta\theta\sigma}
\ee
 Diagonalizing (\ref{eq: 2by2matrix}) is of course trivial
\cite{Yeomans}. Here, however, we seek a method which does not
require knowledge of the characteristic polynomial of the matrix,
so that it can be generalized to replicated transfer matrices with
arbitrary $n$. To this end we introduce the two vectors
$\bu_{0}[x],\bu_{1}[x,\mu]\in\R^2$, parametrized by $x,\mu\in\R$,
and with components
\be
 u_{0}(\sigma;x)=e^{\beta x\sigma} \qquad
 u_{1}(\sigma;x,\mu)=e^{\beta x\sigma}(\sigma-\mu)
 \label{eq:right_candidates}
\ee
 Inserting the candidates (\ref{eq:right_candidates}) into the
eigenvalue equation
$\sum_{\sigma^\prime}T(\sigma,\sigma^\prime;\theta,J)u(\sigma^\prime)=\lambda
u(\sigma)$, and using the general identity
$f(\sigma)=e^{\beta[B+A\sigma]}$ where
$A=\frac{1}{2\beta}\log[f(1)/f(-1)]$ and
$B=\frac{1}{2\beta}\log[f(1)f(-1)]$, for $\sigma\in\{-1,1\}$,
leads to the following eigenvalue equations:
\begin{eqnarray}
\label{eq: eig0}
  e^{\beta B(J,x)+\beta[\theta+A(J,x)]\sigma}&=&\lambda_{0}e^{\beta x\sigma}\qquad
\\ \label{eq: eig1}
  e^{\beta B(J,x)+\beta[\theta+A(J,x)]\sigma}
  A^\prime(J,x)
     \Big(\sigma-\frac{\mu- B^\prime(J,x)}{A^\prime(J,x)}\Big)
  &=&\lambda_{1}e^{\beta x\sigma}(\sigma-\mu)
\end{eqnarray}
where
\begin{eqnarray}
  A(J,x)&=&\frac{1}{\beta}\arctanh[\tanh(\beta J)\tanh(\beta x)]
  \label{eq:define_A}
\\
 B(J,x)&=&\frac{1}{2\beta}\log[
 4\cosh(\beta(J+x))\cosh(\beta(J-x))]
 \label{eq:define_B}
\end{eqnarray}
with partial derivatives
 $A^\prime(J,x)= \partial_x A(J,x)=\frac{1}{2}[\tanh(\beta J+\beta x)+\tanh(\beta J-\beta x)]$ and
 $B^\prime(J,x)=\partial_x B(J,x)=\frac{1}{2}[\tanh(\beta J+\beta x)-\tanh(\beta J-\beta
 x)]$, respectively. We conclude from (\ref{eq: eig0}) that if $x^{*}$ is the solution
of the algebraic equation $x=\theta+A(J,x)$, then $\bu_{0}[x^{*}]$
is an eigenvector with eigenvalue $\lambda_{0}=e^{\beta
B(J,x^{*})}$. This (unique) solution, which can be viewed as the
stable fixed point of the iterative map
$x_{i+1}=\theta+A(J,x_{i})$, is given by
\be
\label{XSTAR}
  x^{*}=\frac{1}{2}(J+\theta)+\frac{1}{2\beta}
    \log\Big[e^{\beta J}\sinh(\beta\theta)+\sqrt{e^{2\beta J}\sinh^{2}(\beta\theta)+e^{-2\beta J}}\Big]
    \label{eq:xstar}
\ee Inserting (\ref{eq:xstar}) into our expression for
$\lambda_{0}$ then reproduces the familiar result for the largest
eigenvalue of the transfer matrix of the Ising chain with uniform
fields and bonds
\be
  \lambda_{0}=e^{\beta B(J,x^{*})}=
    e^{\beta J}\cosh(\beta\theta)+\sqrt{e^{2\beta J}\sinh^{2}(\beta\theta)+e^{-2\beta J}}
\ee Similarly we see that if
$\mu^{*}=\frac{B'(J,x^{*})}{1-A'(J,x^{*})}$, with $x^*$ as defined
before, then also $\bu_{1}[x^{*},\mu^{*}]$ is an eigenvector with
eigenvalue $\lambda_{1}=e^{\beta B(J,x{^*})}A'(J,x^{*})$.
Insertion of (\ref{eq:xstar}) leads to the familiar expression for
the second eigenvalue of (\ref{eq: 2by2matrix}):
\be
  \lambda_{1}=
   e^{\beta J}\cosh(\beta\theta)-\sqrt{e^{2\beta J}\sinh^{2}(\beta\theta)+e^{-2\beta J}}
\ee It turns out that $\mu^{*}$ gives the average magnetization at
each site:
\be
  \mu^{*}=\frac{\tanh(\beta x^{*})[1+\tanh(\beta J)]}
                 {1+\tanh(\beta J)\tanh^{2}(\beta x^{*})}=
    \frac{\sinh(\beta\theta)}{\sqrt{\sinh^{2}(\beta\theta)+e^{-4\beta J}}}=
    \bra\sigma_{i}\ket
\ee Note that our expression for (\ref{eq: 2by2matrix}) is not
symmetric (although one could easily write the partition sum in
terms of a symmetric transfer matrix), hence  we have to
distinguish between left and right eigenvectors; so far only right
eigenvectors have been calculated. We can find the left
eigenvectors $\bv$ via similar ansatz to
(\ref{eq:right_candidates}):
\be
 v_{0}(\sigma;y)=e^{\beta y\sigma} \qquad
 v_{1}(\sigma;y,\nu)=e^{\beta y\sigma}(\sigma-\nu)
 \label{eq:left_candidates}
\ee Insertion into the left eigenvalue equation
$\sum_{\sigma^\prime}v(\sigma^\prime)T(\sigma^\prime,\sigma;\theta,J)=
    \lambda v(\sigma)$
then reveals that the two vectors $\bv_{0}[y^{*}]$ and
$\bv_{1}[y^{*},\nu^{*}]$ are left eigenvectors,  where $y^{*}$ is
the solution of $y^{*}=A(J,y^{*}+\theta)$ and
$\nu^{*}=\frac{B'(J,y^{*}+\theta)}{1-A'(J,y^{*}+\theta)}$. The
associated eigenvalues are $\lambda_{0}=e^{\beta
B(J,y^{*}+\theta)}$and $\lambda_{1}=e^{\beta
B(J,y^{*}+\theta)}A'(J,y^{*}+\theta)$. The fixed point $y^{*}$ of
the map $y_{i+i}=A(J,y_{i}+\theta)$ is again unique, and is given
by:
\be
\label{YSTAR}
  y^{*}=\frac{1}{2}(J-\theta)+\frac{1}{2\beta}
      \log\Big[e^{\beta J}\sinh(\beta\theta)+\sqrt{e^{2\beta J}\sinh^{2}(\beta\theta)+e^{-2\beta J}}\Big]
\ee Obviously  $x^{*}=y^{*}+\theta$, so left and right eigenvalues
are identical and $\nu^{*}=\mu^{*}=\bra\sigma_{i}\ket$.
Furthermore, upon using the simple relation $\tanh(\beta
x^{*}+\beta y^{*})=\bra\sigma_{i}\ket=\mu^{*}=\nu^{*}$
 it is clear that left and right
eigenvectors corresponding to different eigenvalues are
orthogonal:
\begin{eqnarray*}
  \sum_{\sigma}v_{0}(\sigma;y^{*})u_{1}(\sigma;x^{*},\mu^{*})&=&
    2\cosh(\beta x^{*}+\beta y^{*})[\tanh(\beta x^{*}+\beta
    y^{*})-\mu^{*}]=0
\\
  \sum_{\sigma}v_{1}(\sigma;y^{*},\mu^{*})u_{0}(\sigma;x^{*})&=&
    2\cosh(\beta x^{*}+\beta y^{*})[\tanh(\beta x^{*}+\beta
    y^{*})-\nu^{*}]=0
\end{eqnarray*}
 Finally, to normalize our
eigenvectors we require the constants
\begin{eqnarray}
\hspace*{-10mm}
  D_{0}(x^{*},y^{*})&=&\sum_{\sigma}v_{0}(\sigma;y^{*})u_{0}(\sigma;x^{*})=2\cosh(\beta x^{*}+\beta y^{*})
\\
\hspace*{-10mm}
  D_{1}(x^{*},y^{*})&=&\sum_{\sigma}v_{1}(\sigma;y^{*},\mu^{*})u_{1}(\sigma;x^{*},\mu^{*})
       =2\cosh(\beta x^{*}+\beta y^{*})\Big[1-(\mu^{*})^{2}\Big]
\end{eqnarray}

\subsection{Uncoupled replicated chains}

As a intermediate step from the the diagonalization of (\ref{eq:
2by2matrix}) for the simple Ising chain  to diagonalization of
(\ref{rtm_1d}) for disordered chains, let us now inspect
replicated transfer matrices with uncoupled replicas, viz.
(\ref{rtm_1d}) but with $\delta$-distributed bonds and fields:
\be
\label{RS}
  T_{n}(\bsigma,\bsigma^\prime;\theta,J)=
    e^{\beta
    J\sum_{\alpha}\sigma_{\alpha}\sigma_{\alpha}^\prime+\beta\theta\sum_{\alpha}\sigma_{\alpha}}
\ee without an average over $\{\theta,J\}$. This matrix is just
the $n$-fold Kronecker product of (\ref{eq: 2by2matrix}), so its
left- and right eigenvectors are simply (Kronecker) products of
(\ref{eq:left_candidates}) and (\ref{eq:right_candidates}),
respectively. Each eigenvector is characterized by an index set
$\{\rho\}\subseteq\{1,\ldots,n\}$ of size $\rho\in\{0,\ldots,n\}$,
indicating those indices $\alpha$ for which we select
$\bu_{1}[x^{*}]$ as opposed to $\bu_{0}[x^{*}]$ (and similarly for
left eigenvectors), and with
   $\{0\}=\emptyset$. The left- and right eigenvectors of (\ref{RS}) can thus be
written as
\begin{eqnarray}
\label{LEV_RS}
  v_{\{\rho\}}(\bsigma;y^{*},\mu^{*})&=&
    \prod_{\alpha\in\{\rho\}}v_{1}(\sigma_{\alpha};y^{*},\mu^{*})\prod_{\alpha\notin\{\rho\}}v_{0}(\sigma_{\alpha};y^{*})
\\
\label{REV_RS}
  u_{\{\rho\}}(\bsigma;x^{*},\mu^{*})&=&
    \prod_{\alpha\in\{\rho\}}u_{1}(\sigma_{\alpha};x^{*},\mu^{*})\prod_{\alpha\notin\{\rho\}}u_{0}(\sigma_{\alpha};x^{*})
\end{eqnarray}
For each $\rho\in\{0,\ldots,n\}$ there are ${n \choose \rho}$
different index subsets, giving us the required total number of
$2^{n}$ eigenvectors. The associated eigenvalues follow easily,
since here all spin summations factorize over replicas:
\begin{eqnarray*}
  &&\hspace*{-10mm}
  \sum_{\bsigma^\prime}T_{n}(\bsigma,\bsigma^\prime;\theta,J)u_{\{\rho\}}(\bsigma^\prime;x^{*},\mu^{*}) \\
  &&=
    \prod_{\alpha\in\{\rho\}}\sum_{\sigma_{\alpha}^\prime}T(\sigma_{\alpha},\sigma_{\alpha}^\prime;\theta,J)u_{1}(\sigma_{\alpha}^\prime;x^{*},\mu^{*})
    \prod_{\alpha\notin\{\rho\}}\sum_{\sigma_{\alpha}^\prime}T(\sigma_{\alpha},\sigma_{\alpha}^\prime;\theta,J)u_{0}(\sigma_{\alpha}^\prime;x^{*})  \\
  &&=\lambda_{1}^{\rho}\lambda_{0}^{n-\rho}\prod_{a\in\{\rho\}}u_{1}(\sigma_{a}';x^{*},\mu^{*})
            \prod_{\alpha\notin\{\rho\}}u_{0}(\sigma_{\alpha}^\prime;x^{*})
  =\lambda_{1}^{\rho}\lambda_{0}^{n-\rho}u_{\{\rho\}}(\bsigma;x^{*},\mu^{*})
\end{eqnarray*}
Hence (\ref{RS}) has $n+1$ different eigenvalues
$\lambda_{\rho}(n)=\lambda_{1}^{\rho}\lambda_{0}^{n-\rho}$, each
with multiplicity ${n \choose \rho}$. Since
$\lambda_{0}>\lambda_{1}$, we also have the ordering relation
$\lambda_{0}(n)>\lambda_{1}(n)>\ldots>\lambda_{n}(n)$. We can
furthermore see that right and left eigenvectors satisfy the
orthogonality relations
\begin{eqnarray}
  \bv_{\{\rho^\prime\}}[y^{*},\mu^{*}]\cdot\bu_{\{\rho\}}[x^{*},\mu^{*}]=
    D_{\rho}(x^{*},y^{*})\delta_{\rho\rho^\prime}\prod_{k=1}^{\rho}\delta_{\alpha_{k}\alpha^\prime_{k}}
  \label{eq:innerprod}  \\
  D_{\rho}(x^{*},y^{*})=
    2\cosh^{n}(\beta x^{*}+\beta y^{*})\left[1-(\mu^{*})^{2}\right]^{\rho}
\end{eqnarray}
where $\{\rho\}=\{\alpha_1,\ldots,\alpha_\rho\}$ and
$\{\rho^\prime\}=\{\alpha^\prime_1,\ldots,\alpha^\prime_{\rho^\prime}\}$,
and where the factor $\prod_{k=1}^\rho
\delta_{\alpha_{k}\alpha^\prime_{k}}$ in (\ref{eq:innerprod}) is
defined as unity for $\rho=0$.

\subsection{Diagonalization for the disordered Ising chain}
\label{sec:eigenvectors}

We now turn the real problem: the diagonalization of
(\ref{rtm_1d}), which can also be written as
$\bT_{n}=\bra\bT_{n}[\theta,J]\ket_{\theta,J}$. Clearly $\bT_{n}$
shares many properties with $\bT_{n}[\theta,J]$, e.g. invariance
under all permutations $\pi$ of the permutation group $S_n$ acting
on the indices $\{1,\ldots,n\}$:
\be
\label{eq: RS_property}
T_{n}(\pi(\bsigma),\pi(\bsigma^\prime))=T_{n}(\bsigma,\bsigma^\prime)
\qquad
  \textrm{for every}\quad \pi\in S_{n}
\ee It follows that if $\bu$ is an eigenvector of $\bT_{n}$ with
eigenvalue $\lambda$, then so is $\bD_{\pi}\bu$ for any $\pi\in
S_{n}$ where $\bD_{\pi}$ denotes the matrix representation of
$\pi$, i.e.
$D_\pi(\bsigma,\bsigma^\prime)=\delta_{\pi(\bsigma),\bsigma^\prime}$.
In the uncoupled case (\ref{RS}) one observes that
 $\bD_{\pi}\bu_{\{a_{1},\ldots,a_{\rho}\}}=\bu_{\{\pi(a_{1}),\ldots,\pi(a_{\rho})\}}$
for every $\pi\in S_{n}$; we will make the ansatz that this also
holds for the eigenvectors of (\ref{eq: RS_property}). The result
is again a spectrum of $n+1$ different eigenvalues
$\lambda_{\rho}(n)$ with $\rho=0,1,\ldots,n$,  with multiplicity
${n \choose \rho}$ each. These statements reproduce the results in
\cite{weigt-monasson96}, which were derived using the irreducible
representations of the replica permutation group. Here we are now
being led  to the following general ansatz for the right- and left
 eigenvectors of $\bT_{n}$:
\begin{eqnarray}
\label{eq: right}
  u_{\{\rho\}}(\bsigma;P_{\rho})&=&
    \int\!\mathrm{d}x\mathrm{d}\mu~P_{\rho}(x,\mu\vert n)~e^{\beta x\sum_{\alpha=1}^{n}\sigma_{\alpha}}
       \prod_{\alpha\in\{\rho\}}(\sigma_{\alpha}-\mu)
\\ \label{eq: left}
    v_{\{\rho\}}(\bsigma;Q_{\rho})&=&
    \int\!\mathrm{d}y\mathrm{d}\nu~Q_{\rho}(y,\nu\vert n)~e^{\beta y\sum_{\alpha=1}^{n}\sigma_{\alpha}}
       \prod_{\alpha\in\{\rho\}}(\sigma_{\alpha}-\nu)
\end{eqnarray}
with $P_{\rho}$ and $Q_{\rho}$ denoting functions to be
determined,  by inserting (\ref{eq: right}) into the right
eigenvalue equation
$\bT_{n}\bu_{\{\rho\}}[P_{\rho}]=\lambda_{\rho}(n)\bu_{\{\rho\}}[P_{\rho}]$,
and (\ref{eq: left}) into the left eigenvalue equation
$\bv_{\{\rho\}}[Q_{\rho}]\bT_{n}=\lambda_{\rho}(n)\bv_{\{\rho\}}[Q_{\rho}]$,
respectively. Working out the first equation gives, with the
definitions (\ref{eq:define_A},\ref{eq:define_B}):
\begin{eqnarray*}
\hspace*{-5mm}\sum_{\bsigma^\prime}T_{n}(\bsigma,\bsigma^\prime)u_{\{\rho\}}(\bsigma^\prime;P_{\rho})
&=&\int\!\mathrm{d}x^\prime\mathrm{d}\mu^\prime~P_{\rho}(x^\prime,\mu^\prime\vert
n) \bigg\bra\prod_{\alpha\notin\{\rho\}}e^{\beta
B(J,x^\prime)+\beta[\theta+A(J,x^\prime)]\sigma_{\alpha}}
    \\
  &&\qquad\qquad\hspace*{-40mm}
  \times
    \prod_{\alpha\in\{\rho\}}e^{\beta
    B(J,x^\prime)+\beta[\theta+A(J,x\prime)]\sigma_{\alpha}}
    A^\prime(J,x^\prime)
       \Big(\sigma_{\alpha}-\frac{\mu^\prime-B^\prime(J,x^\prime)}{A^\prime(J,x^\prime)}\Big)\bigg\ket_{J,\theta}
\end{eqnarray*}
Upon inserting suitable integrals over $\delta-$functions, viz.
$1=\int\!\mathrm{d}x~\delta[x-\theta-A(J,x^\prime)]$ and
  $1=\int\!\mathrm{d}\mu~\delta[\mu-\frac{\mu^\prime-B^\prime(J,x^\prime)}{A^\prime(J,x^\prime)}]$,
we then find our right  eigenvalue equation taking the form
\small
\begin{eqnarray*}
\hspace*{-25mm} \int\!\mathrm{d}x\mathrm{d}\mu~
              \bigg[\int\!\mathrm{d}x^\prime\mathrm{d}\mu^\prime~P_{\rho}(x^\prime\!,\mu^\prime\vert n)\Big\bra
                      e^{n\beta B(J,x^\prime)}[A^\prime(J,x^\prime)]^{\rho}
                       \delta[x-\theta-A(J,x^\prime)]\delta[\mu-\frac{\mu^\prime\!-B^\prime(J,x^\prime)}{A^\prime(J,x^\prime)}]
              \Big\ket_{J,\theta}\bigg]  \\
          \hspace*{30mm}\times \bigg[
e^{\beta x\sum_{a=1}^{n}\sigma_{a}}
       \prod_{a\in\{\rho\}}(\sigma_{a}-\mu) \bigg]\\
  =
    \lambda_{\rho}(n)~\int\!\mathrm{d}x\mathrm{d}\mu~P_{\rho}(x,\mu\vert n)~\bigg[ e^{\beta x\sum_{a=1}^{n}\sigma_{a}}
       \prod_{a\in\{\rho\}}(\sigma_{a}-\mu)\bigg]
\end{eqnarray*}
\normalsize
We conclude from this that the function $P_{\rho}$ must satisfy
the following  eigenvalue equation:
\be
\label{eq:eigP}
 \int\!\mathrm{d}x^\prime\mathrm{d}\mu^\prime~ \Lambda^{(P)}_{\rho}(x,\mu,x^\prime\!,\mu^\prime\vert n)~P_{\rho}(x^\prime\!,\mu^\prime\vert n)
= \lambda_{\rho}(n)~P_{\rho}(x,\mu\vert n)
 \ee
  with the  kernel
\begin{eqnarray}
  \Lambda^{(P)}_{\rho}(x,\mu,x^\prime\!,\mu^\prime\vert n)&=&
   \Big\bra e^{n\beta B(J,x^\prime)}[A^\prime(J,x^\prime)]^{\rho} \nonumber
   \\
   && \times~\delta[x-\theta-A(J,x^\prime)]
   \delta[\mu-\frac{\mu^\prime\!-B^\prime(J,x^\prime)}{A^\prime(J,x^\prime)}]
              \Big\ket_{J,\theta}
              \label{eq:kerP}
\end{eqnarray}
Upon repeating the above procedure also for the left eigenvectors
(\ref{eq: left}) we find a similar eigenvalue problem for the
functions $Q_{\rho}$, but now with a different kernel
$\Lambda^{(Q)}_{\rho}$:
\be
\label{eq:eigQ}
 \int\!\mathrm{d}y^\prime\mathrm{d}\nu^\prime~
    \Lambda^{(Q)}_{\rho}(y,\nu,y^\prime,\nu^\prime\vert n)~Q_{\rho}(y^\prime\!,\nu^\prime\vert n)=
  \lambda_{\rho}(n)~Q_{\rho}(y,\nu\vert n)
\ee
\begin{eqnarray}
  \Lambda^{(Q)}_{\rho}(y,\nu,y^\prime\!,\nu^\prime\vert n)&=&
   \Big\bra e^{n\beta
   B(J,y^\prime\!+\theta)}[A^\prime(J,y^\prime+\theta)]^{\rho}
   \nonumber
   \\
   && \times~
\delta[y-A(J,y^\prime\!+\theta)]\delta[\nu-\frac{\nu^\prime\!-B^\prime(J,y^\prime\!+\theta)}{A^\prime(J,y^\prime\!+\theta)}]
              \Big\ket_{J,\theta}
              \label{eq:kerQ}
\end{eqnarray}
We have now transformed the problem of diagonalizing the
$2^n\times 2^n$ replicated transfer matrix (\ref{rtm_1d}) into a
problem involving integral operators
(\ref{eq:kerP},\ref{eq:kerQ}), where the limit $n\to 0$ can be
taken.

We note that, at least for the purpose at finding the eigenvalues
$\lambda_{\rho}(n)$, the two eigenvalue problems
(\ref{eq:eigP},\ref{eq:eigQ}) can be integrated over $\mu$ and
$\nu$, respectively, and replaced by a simpler eigenvalue problem
for the two single-argument functions $\Phi_{\rho}(x\vert
n)=\int\!\mathrm{d}\mu~P_{\rho}(x,\mu\vert n)$ and
$\Psi_{\rho}(y\vert n)=\int\!\mathrm{d}\nu~Q_{\rho}(y,\nu\vert
n)$:
\begin{eqnarray}
\int\!\mathrm{d}x^\prime~\Lambda_\rho^{(P)}(x,x^\prime|n)
~\Phi_\rho(x^\prime|n)&=& \lambda_{\rho}(n)~\Phi_{\rho}(x|n)
\\
\int\!\mathrm{d}y^\prime~\Lambda_\rho^{(Q)}(y,y^\prime|n)
~\Psi_\rho(y^\prime|n)&=& \lambda_{\rho}(n)~\Psi_{\rho}(y|n)
\end{eqnarray}
with
\begin{eqnarray}
  \Lambda^{(P)}_{\rho}(x,x^\prime|n)&=&
   \Big\bra e^{n\beta B(J,x^\prime)}[A^\prime(J,x^\prime)]^{\rho} \delta[x-\theta-A(J,x^\prime)]
              \Big\ket_{J,\theta}
              \label{eq: kerPsimple}
\\
  \Lambda^{(Q)}_{\rho}(y,y^\prime|n)&=&
   \Big\bra e^{n\beta
   B(J,y^\prime\!+\theta)}[A^\prime(J,y^\prime+\theta)]^{\rho}\delta[y-A(J,y^\prime\!+\theta)]
              \Big\ket_{J,\theta}
              \label{eq: kerQsimple}
\end{eqnarray}
Once we know the functions $P_{\rho}$ and $Q_{\rho}$, the form of
the kernels (\ref{eq: kerPsimple},\ref{eq: kerQsimple}) enables us
to integrate (\ref{eq:eigP}) and (\ref{eq:eigQ}) over $x$ and $y$,
and obtain relatively expressions for the corresponding
eigenvalues:
\begin{eqnarray}
  \lambda_{\rho}(n)&=&
   \frac{\int\!\mathrm{d}x~\Phi_{\rho}(x\vert n)
      \Big\bra e^{n\beta B(J,x)}[A^\prime(J,x)]^{\rho}\Big\ket_{J}}
        {\int\!\mathrm{d}x~\Phi_{\rho}(x\vert n)}
  \label{eq: eigenvalues} \\
  \lambda_{\rho}(n) &=&
   \frac{\int\!\mathrm{d}y~\Psi_{\rho}(y\vert n)
      \Big\bra e^{n\beta B(J,y+\theta)}[A^\prime(J,y+\theta)]^{\rho}\Big\ket_{J,\theta}}
        {\int\!\mathrm{d}y~\Psi_{\rho}(y\vert n)}
        \label{eq: eigenvalues2}
\end{eqnarray}

Let us quickly inspect special cases. We see that for
$\delta-$distributed fields and bonds the integral equations
(\ref{eq:eigP},\ref{eq:eigQ}) admit the expected solutions
$P_{\rho}(x,\mu\vert n)=\delta(x-x^{*})\delta(\mu-\mu^{*})$ and
$Q_{\rho}(y,\nu\vert n)=\delta(y-y^{*})\delta(\nu-\mu^{*})$, the
eigenvectors (\ref{eq: right},\ref{eq: left}) reduce to the
eigenvectors of (\ref{RS}), and the eigenvalues become
$\lambda_{\rho}(n)=\lambda_{1}^{\rho}\lambda_{0}^{n-\rho}$, as
they should. Also the special case of a chain without external
fields, i.e. $p(J,\theta)=p(J)\delta(\theta)$, can be easily
solved analytically. Here $A(J,0)=B^\prime(J,0)=0$ and
$A^\prime(J,0)=\tanh(\beta J)$ for every $J$, which enables us to
verify that (\ref{eq:eigP},\ref{eq:eigQ}) have the trivial
solutions $P_{\rho}(x,\mu\vert n)=\delta(x)\delta(\mu)$ and
$Q_{\rho}(y,\nu\vert n)=\delta(y)\delta(\nu)$. Hence the
eigenvectors become \bd
  u_{\{\rho\}}(\bsigma)=v_{\{\rho\}}(\bsigma)=\prod_{\alpha\in\{\rho\}}\sigma_{\alpha}
\ed They satisfy
$\bv_{\rho}[Q]\cdot\bu_{\rho^\prime}[P]=2^{n}\delta_{\rho\rho^\prime}\prod_{k=1}^{\rho}\delta_{\alpha_{k}\alpha^\prime_{k}}$.
These eigenvectors are in fact common to all matrices of the form
$T(\bsigma,\bsigma^\prime)=T(\bsigma\cdot\bsigma^\prime)$
\cite{van-hemmen87}, and our replicated transfer matrix  falls in
this category when the external fields are zero. The eigenvalues
are given by $\lambda_{\rho}(n)=\big\bra [2\cosh(\beta
J)]^{n}\tanh^{\rho}(\beta J)\big\ket_{J}$, and it is clear that
the largest corresponds to $\rho=0$.

\subsection{Properties of the kernel eigenvalue problems for $n\to 0$}

Let us consider in more detail the $n\to 0$ limits of the
eigenvalue problems (\ref{eq:kerP}) and (\ref{eq:kerQ}). We first
turn to $\rho=0$. The eigenvectors corresponding to eigenvalue
$\lambda_{0}(0)$ do not depend on $\{\nu,\mu\}$, so upon writing
simply $\Phi_0(x|0)=\Phi(x)$ and $\Psi_0(y|0)=\Psi(y)$ we obtain
for $\rho=0$:
\begin{eqnarray}
\label{eq: PHI}
\int\mathrm{d}x^\prime~\Phi(x^\prime)\Big\bra\delta[x-\theta-A(J,x^\prime)]\Big\ket_{J,\theta}&=&
  \lambda_{0}(0)~\Phi(x)
\\
 \label{eq: PSI}
 \int\mathrm{d}y^\prime~\Psi(y^\prime)\Big\bra\delta[y-A(J,y^\prime+\theta)]\Big\ket_{J,\theta}&=&
  \lambda_{0}(0)~\Psi(y)
\end{eqnarray}
If we assume that $\int\!\mathrm{d}x~\Phi(x)\neq 0$ and
$\int\!\mathrm{d}y~\Psi(y)\neq 0$, then integration of (\ref{eq:
PHI},\ref{eq: PSI}) over $x$ and $y$, respectively, gives us in
both equations  $\lambda_{0}(0)=1$. This, in turn, implies that
$\Phi(x)$ and $\Psi(y)$ are the stationary distributions of the
two random maps \bd
  x_{i+1}=\theta_{i}+A(J_{i},x_{i})\qquad\qquad
  y_{i+1}=A(J_{i},y_{i}+\theta_{i})
\ed These maps describe the propagation of the fields $x$ and $y$
along the chain. The two distributions are connected via the
following equations,
\be
\label{eq: PHI-PSI}
  \Phi(x)=\int\!\mathrm{d}y~\Psi(y)\bra\delta[x-\theta-y]\ket_{\theta}\quad\quad
  \Psi(y)=\int\!\mathrm{d}x~\Phi(x)\bra\delta[y-A(J,x)]\ket_{J}
\ee which can be verified upon substituting into (\ref{eq:
PHI},\ref{eq: PSI}), using $\lambda_{0}(0)=1$. \vsp

The case $\rho>0$ is more complicated. Here we find the $n\to 0$
eigenvalue problems
\small
\begin{eqnarray}
 \int\!\mathrm{d}x^\prime\mathrm{d}\mu^\prime
    \Big\bra [A^\prime(J,x^\prime)]^{\rho}\delta[x-\theta-A(J,x^\prime)]
       \delta\Big[\mu-\frac{\mu^\prime\!-B^\prime(J,x^\prime)}{A^\prime(J,x^\prime)}\Big]
   \Big\ket_{J,\theta} P_{\rho}(x^\prime\!,\mu^\prime\vert
   0)\nonumber
   \\
  \hspace*{45mm} =\lambda_{\rho}(0)~P_{\rho}(x,\mu\vert 0)
  \label{eq: P}
\\
 \int\!\mathrm{d}y^\prime\mathrm{d}\nu^\prime
    \Big\bra[A^\prime(J,y^\prime\!+\theta)]^{\rho}\delta[y-A(J,y^\prime\!+\theta)]
      \delta\Big[\mu-\frac{\mu^\prime\!-B^\prime(J,y^\prime\!+\theta)}{A^\prime(J,y^\prime\!+\theta)}\Big]
    \Big\ket_{J,\theta} Q_{\rho}(y^\prime\!,\nu^\prime\vert 0)
 \hspace*{-5mm}  \nonumber \\
    \hspace*{45mm}
  =\lambda_{\rho}(0)~Q_{\rho}(y,\nu\vert 0)
  \label{eq: Q}
\end{eqnarray}
\normalsize
As for $\rho=0$ we can show that these equations admit solutions
$P_{\rho}$ and $Q_{\rho}$ which can be interpreted as probability
densities. The difference with
$\rho=0$, where these distributions are the stationary measures of
the random maps of the propagated fields $\{x,y\}$, is that here
the quantities which are propagated are the distributions
themselves, via deterministic but nonlinear functional maps: \bd
  P_{\rho,i+1}=\mathcal{A}_{P,\rho}(P_{\rho,i}) \qquad\qquad
  Q_{\rho,i+1}=\mathcal{A}_{Q,\rho}(Q_{\rho,i})
\ed
where
\begin{eqnarray}
[\mathcal{A}_{P,\rho}(P)](x,\mu)&=&
    \int\!\mathrm{d}x^\prime\mathrm{d}\mu^\prime~ \bigg\bra
    \frac{P(x^\prime,\mu^\prime)[A^\prime(J,x^\prime)]^{\rho}}
         {\int\!\mathrm{d}x^\pprime\mathrm{d}\mu^\pprime P(x^\pprime,\mu^\pprime)\bra[A^\prime(J^\pprime\!,x^\pprime)]^{\rho}\ket_{J^\pprime}}
  \nonumber \\
  &&\qquad\times
    \delta[x-\theta-A(J,x^\prime)]\delta[\mu-\frac{\mu^\prime\!-B^\prime(J,x^\prime)}{A^\prime(J,x^\prime)}]
    \bigg\ket_{J,\theta}
    \label{eq:fmapP}
\\[0mm]
[\mathcal{A}_{Q,\rho}(Q)](y,\nu)&=&
    \int\!\mathrm{d}y^\prime\mathrm{d}\nu^\prime ~\bigg\bra
    \frac{Q(y^\prime\!,\nu^\prime)[A^\prime(J,y^\prime\!+\theta)]^{\rho}}
         {\int\!\mathrm{d}y^\pprime\mathrm{d}\nu^\pprime Q(y^\pprime,\nu^\pprime)
          \bra[A^\prime(J^\pprime,y^\pprime\!+\theta^\pprime)]^{\rho}\ket_{J^\pprime\!,\theta^\pprime}}
  \nonumber \\
  &&\qquad\times
    \delta[y-A(J,y^\prime\!+\theta)]\delta[\nu-\frac{\nu^\prime\!-B^\prime(J,y^\prime\!+\theta)}{A^\prime(J,y^\prime\!+\theta)}]
    \bigg\ket_{J,\theta}
    \label{eq:fmapQ}
\end{eqnarray}
We see that the defining properties of a probability density, viz.
non-negativity and normalization, are preserved by both functional
maps. Hence we may indeed view the eigenvalue problems (\ref{eq:
P},\ref{eq: Q}) as the fixed point equations of the functional
maps (\ref{eq:fmapP},\ref{eq:fmapQ}). The eigenvalues are:
\be
  \lambda_{\rho}(0)=\int\mathrm{d}x\,\Phi_{\rho}(x\vert 0)
   \bra[A'(J,x)]^{\rho}\ket_{J}=
   \int\mathrm{d}y\,\Psi_{\rho}(y\vert 0)
   \bra[A'(J,y+\theta)]^{\rho}\ket_{\theta,J}
\ee where $\Phi_{\rho}$ and $\Psi_{\rho}$ are as before the
marginals of $P_{\rho}$ and $Q_{\rho}$. Moreover, using the
property $A(J,x)<1$ for every $J,x$ we obtain
$\lambda_{\rho}(0)<\lambda_{0}(0)=1$ for every $\rho>1$. We may
also generalize equations (\ref{eq: PHI-PSI}) which give the
relation between the solutions of the two $\rho=0$ eigenvalue
problems. It straightforward to check by substitution into
(\ref{eq: P},\ref{eq: Q}) that for $\rho>1$ we have:
\small
\begin{eqnarray}
  P_{\rho}(x,\mu\vert 0)&=&
    \int\mathrm{d}y\mathrm{d}\nu\,Q_{\rho}(y,\nu\vert 0)
    \bra\delta(x-\theta-y)\ket_{\theta}\delta(\mu-\nu)
\\
  Q_{\rho}(y,\nu\vert 0)&=&
  \frac{\int\mathrm{d}x\mathrm{d}\mu\,P_{\rho}(x,\mu\vert 0)
        \bra[A'(J,x)]^{\rho}\delta[y-A(J,x)]\delta[\nu-\frac{\mu-B'(J,x)}{A'(J,x)}]\ket_{J}}
       {\int\mathrm{d}x\,\Phi_{\rho}(x\vert 0)\bra[A'(J,x)]^{\rho}\ket_{J}}
\end{eqnarray}
\normalsize

\subsection{Spectral decompositions}

Standard linear algebra guarantees that left- and right
eigenvectors corresponding to different eigenvectors are
orthogonal. This, given that our eigenvalues $\lambda_\rho(n)$
depend only on the size $\rho$ of the index sets
 we know that
\begin{eqnarray}
\rho\neq \rho^\prime: &~~~& \sum_{\bsigma}
  u_{\{\rho\}}(\bsigma;P_{\rho})
    v_{\{\rho^\prime\}}(\bsigma;Q_{\rho^\prime})=0
 \label{eq:orthogonality}
\end{eqnarray}
It follows that we may always use the decomposition
\begin{eqnarray}
\label{eq:spec_dec}
  T_{n}(\bsigma,\bsigma^\prime)&=&
\sum_{\rho=0}^{n}\lambda_\rho(n)
U_n^{(\rho)}(\bsigma,\bsigma^\prime)
\end{eqnarray}
in which the matrices $\bU_n^{(\rho)}$ are projection matrices,
each formed of linear combinations of
$\lambda_\rho(n)$-eigenvectors and each acting only in one of the
orthogonal eigenspaces.  We note that also
$\bT^k_n=\sum_{\rho=0}^n \lambda_\rho^k(n)\bU_n^{(\rho)}$ for any
integer $k>0$, and that the trace of a projection operator reduces
to the dimension of the space which it projects, i.e.
$\tr(\bU_n^{(\rho)})={n \choose \rho}$. Since the dimensions of
both the $\lambda_0(n)$ and the $\lambda_n(n)$ eigenspaces are
one, the corresponding eigenvectors are pairwise orthogonal and
orthogonal to all other eigenvectors, and therefore
\be
U_n^{(0)}(\bsigma,\bsigma^\prime)=
 \frac{u_{\{0\}}(\bsigma)v_{\{0\}}(\bsigma^\prime)}{D_{0}(n)}
~~~~~~~~ U_n^{(n)}(\bsigma,\bsigma^\prime)=
 \frac{u_{\{n\}}(\bsigma)v_{\{n\}}(\bsigma^\prime)}{D_{n}(n)}
\ee
with
\begin{eqnarray}
  D_{\rho}(n)&=&
     \sum_{\bsigma}v_{\{\rho\}}(\bsigma)u_{\{\rho\}}(\bsigma) \nonumber \\
  &=&\int\!\mathrm{d}x\mathrm{d}\mu~P_{\rho}(x,\mu\vert n)
                \int\!\mathrm{d}y\mathrm{d}\nu~Q_{\rho}(y,\nu\vert n) \nonumber \\
   &&\qquad\times~
    \big[2\cosh(\beta x+\beta y)\big]^{n}
    \Big[1+\mu\nu-\tanh(\beta x+\beta y)[\mu+\nu]\Big]^{\rho}
    \label{eq:expressionforD}
\end{eqnarray}
We note that $\lim_{n\to 0}D_{0}(n)=1$.
 Expression (\ref{eq:spec_dec}) will prove useful in
calculating observables such as magnetizations and correlation
functions. If also within each eigenspace characterized by an
index set size $1\leq \rho\leq n-1$ the eigenvectors would be
orthogonal (as in chains without disorder, or in the random bond
chain without external fields), then we would have
$U_n^{(\rho)}(\bsigma,\bsigma^\prime)=\sum_{\{\rho\}}u_{\{\rho\}}(\bsigma)v_{\{\rho\}}(\bsigma^\prime)/D_{\rho}(n)$
for {\em all} $\rho$, and hence
\begin{eqnarray}
\label{eq:spec_dec_otho}
  T_{n}(\bsigma,\bsigma^\prime)&=&\sum_{\rho=0}^{n}\lambda_{\rho}(n)
   \sum_{\{\rho\}}\frac{u_{\{\rho\}}(\bsigma)v_{\{\rho\}}(\bsigma^\prime)}{D_{\rho}(n)}
\end{eqnarray}

\section{Applications of the theory: the random field Ising model}
\label{sec:rfim}

As a benchmark test, let us first calculate the free energy and
various observables for the random field Ising chain
(\ref{eq:1DH}) with nearest neighbour bonds of strength $J_{0}$.

\subsection{The free energy}

We recall that the free energy is given by: $\bar{f}=-\lim_{n\to
0}\frac{1}{n}\lim_{N\to\infty}\frac{1}{\beta N}
           \log\tr(\bT_{n}^{N})
$,
where $\bT_{n}$ is the replicated transfer matrix (\ref{rtm_1d}). Assuming
that the largest eigenvalue is $\lambda_{0}(n)$, we may write the trace as:
\be
  \tr(\bT_{n}^{N})=\sum_{\rho=0}^{n}
      [\lambda_{\rho}(n)]^N \tr(\bU_n^{(\rho)})=
\lambda_{0}^{N}(n)
    \Big[1+\sum_{\rho=1}^{n}{n \choose \rho}
      \left(\frac{\lambda_{\rho}(n)}{\lambda_{0}(n)}\right)^{\!N}\Big]
\ee
 Since
$\lim_{N\to\infty}(\lambda_{\rho<0}(n)/\lambda_{0}(n))^{N}= 0$,
only the contribution of the largest eigenvalue survives, so that,
upon writing $\lambda_0(n)=1+\lambda n+\order(n^2)$ (for we had
already established that $\lambda_0(0)=1$):
 \be
  \bar{f}= -\frac{1}{\beta}\lim_{n\to 0}\frac{1}{n}\log\lambda_{0}(n)
         = -\frac{1}{\beta}\lim_{n\to 0}\frac{1}{n}\log[1+n\lambda+\order(n^{2})]
         = -\frac{\lambda}{\beta}
         \label{eq:rfic1}
\ee The $\order(n)$ contribution $\lambda$ to $\lambda_{0}(n)$ can
be found upon expanding (\ref{eq: eigenvalues}) for small $n$, and
is found to be $\lambda=\beta\int\!\mathrm{d}x~\Phi(x)B(J_{0},x)$.
Insertion into (\ref{eq:rfic1}) gives us
\be
  \bar{f}= -\frac{1}{2\beta}\int\!\mathrm{d}x~\Phi(x)
    \log 4\cosh(\beta(J_{0}+x))\cosh(\beta(J_{0}-x))
\ee This expression can be converted into a form  more familiar
from  the one dimensional random systems literature
\cite{weigt-monasson96,BrandtGross78,BruinsmaAeppli83}.  If we
define a new random variable $\tilde{x}$ and an associated density
$\tilde{\Phi}(\tilde{x})$ via
 $\tilde{\Phi}(\tilde{x})=\int\!\mathrm{d}x~\Phi(x)
     \delta[\tilde{x}-e^{2\beta x}]$,
we find after some straightforward manipulations that
\be
  \bar{f}=\bra\theta\ket_{\theta}-
    \frac{1}{\beta}\int\!\mathrm{d}\tilde{x}~\tilde{\Phi}(\tilde{x})
    \log[e^{\beta J_{0}}+\tilde{x}e^{-\beta J_{0}}]
    \label{eq:rfic2}
\ee
 where
 \be
  \tilde{\Phi}(\tilde{x})=\int\!\mathrm{d}\tilde{x}^\prime~\tilde{\Phi}(\tilde{x}^\prime)
    \bigg\bra\delta\left[\tilde{x}-e^{2\beta\theta}
      \frac{e^{-\beta J_{0}}+\tilde{x}^\prime e^{\beta J_{0}}}{e^{\beta J_{0}}+\tilde{x}^\prime e^{-\beta J_{0}}}
      \right]\bigg\ket_{\theta}
\ee
 The resulting (correct) expression  (\ref{eq:rfic2}) for the free energy justifies {\em a posteriori}  our
 assumption that
$\lambda_{0}(n)$ as generally the largest eigenvalue,  and
confirms that our ansatz for the associated right- and left
eigenvectors, which are seen themselves to be replica symmetric
(i.e. $u_{0}(\pi(\bsigma))=u_{0}(\bsigma)$ and
$v_{0}(\pi(\bsigma))=v_{0}(\bsigma)$ for every permutation $\pi\in
S_{n}$), was correct.

\subsection{Single site expectation values and their powers}

Let us next show how single-site
 observables of the form $\overline{\bra\sigma_{i}\ket^{\rho}}$ (integer $\rho$),
with brackets denoting a thermal average over the Boltzman measure
and $\overline{\cdots}$ denoting averaging over the disorder, can
also be calculated. We use the following replica identity
\begin{eqnarray}
  \overline{\bra\sigma_{i}\ket^{\rho}}&=&
    \lim_{n\to 0}\overline{\Big[\sum_{\bsigma}\sigma_{i}e^{-\beta H(\bsigma)}\Big]^{\rho}
       \Big[\sum_{\bsigma}e^{-\beta H(\bsigma)}]^{n-\rho}}  \nonumber\\
       &=&\lim_{n\to 0}
     \sum_{\{\bsigma\}}\sigma_{i}^{\alpha_{1}}\ldots\sigma_{i}^{\alpha_{\rho}}
       \overline{\prod_{\alpha=1}^{n}e^{-\beta H(\bsigma^{\alpha})}}
       \label{eq:spin_averages}
\end{eqnarray}
and define the diagonal $2^{n}\times 2^{n}$ matrix
$\bS_{\{\rho\}}$ with entries
\be
  S_{\{\rho\}}(\bsigma,\bsigma^\prime)=\delta_{\bsigma,\bsigma^\prime}
    \prod_{\alpha\in\{\rho\}}\sigma_{\alpha}
\ee
 Upon using the replicated transfer matrix (\ref{rtm_1d}) to
 evaluate (\ref{eq:spin_averages}), and upon dividing (\ref{eq:spin_averages})
 by $1=\lim_{n\to 0}\overline{Z^{n}}=\lim_{n\to 0}\tr(\bT^{N})$,
 expression  (\ref{eq:spin_averages}) can be written in the form
\bd
  \overline{\bra\sigma_{i}\ket^{\rho}}=
    \lim_{n\to 0}\frac{\tr(\bS_{\{\rho\}}\bT^{N})}{\tr(\bT^{N})}
    \label{eq:spin_averages2}
\ed For large $N$ our spectral decomposition (\ref{eq:spec_dec})
now gives us
\begin{eqnarray}
  \overline{\bra\sigma_{i}\ket^{\rho}}&=&
    \lim_{n\to 0}\lim_{N\to\infty}
    \frac{\tr(\bS_{\{\rho\}}\bU_n^{(0)})+ \sum_{\rho^\prime=1}^n [\lambda_{\rho^\prime}(n)/\lambda_0(n)]^N \tr(\bS_{\{\rho\}}\bU_n^{(\rho^\prime)})}
{1+ \sum_{\rho^\prime=1}^n
[\lambda_{\rho^\prime}(n)/\lambda_0(n)]^N } \nonumber
\\
&=&
 \lim_{n\to 0}\tr(\bS_{\{\rho\}}\bU_n^{(0)})\nonumber
 \\
 &=&
\lim_{n\to 0}D_{0}^{-1}(n)
   \sum_{\bsigma}v_{\{0\}}(\bsigma)u_{\{0\}}(\bsigma)  \prod_{\alpha\in\{\rho\}}\sigma_{\alpha} \nonumber \\
  &=&\lim_{n\to 0}\int\mathrm{d}x\mathrm{d}y\,\Phi_{0}(x\vert n)\Psi_{0}(y\vert n)
     \left[2\cosh(\beta x+\beta y)\right]^{n}\tanh^{\rho}(\beta x+\beta y)
\end{eqnarray}
We note that the dependence on the particular realization of the
index set $\{\rho\}$ has disappeared, as it should, leaving only a
dependence on the size $\rho$ of this set. We may now take the
limit $n\to 0$, and find our transparent and appealing final
result
\be
\label{eq:magn}
  \overline{\bra\sigma_{i}\ket^{\rho}}=
    \int\!\mathrm{d}x\mathrm{d}y~\Phi(x)\Psi(y)~\tanh^{\rho}(\beta x+\beta y)
\ee

\subsection{Multiple-site observables}

Finally we apply our methods to the evaluation of
disorder-averaged powers of two-spin correlations, of the form
$\overline{\bra\sigma_{i}\sigma_{j}\ket^{\rho}}$ with integer
$\rho$. We choose $j>i$ and start from the identity
\begin{eqnarray}
  \overline{\bra\sigma_{i}\sigma_{j}\ket^{\rho}}&=&
    \lim_{n\to 0}\overline{
       \Big[\sum_{\bsigma}\sigma_{i}\sigma_{j}e^{-\beta H(\bsigma)}\Big]^{\rho}
       \Big[\sum_{\bsigma}e^{-\beta H(\bsigma)}]^{n-\rho}} \nonumber \\
  &=&\lim_{n\to 0}
     \sum_{\{\bsigma\}}\sigma_{i}^{\alpha_{1}}\sigma_{j}^{\alpha_{1}}
                       \ldots\sigma_{i}^{\alpha_{\rho}}\sigma_{j}^{\alpha_{\rho}}
       \overline{\prod_{\alpha=1}^{n}e^{-\beta H(\bsigma^{\alpha})}}\nonumber \\
  &=&\lim_{n\to 0}
     \frac{\tr(\bS_{\{\rho\}}\bT^{j-i}\bS_{\{\rho\}}\bT^{N-j+i})}{\tr(\bT^{N})}
\end{eqnarray}
Our spectral decomposition (\ref{eq:spec_dec}), together with
$\lambda_0(n)=1$, enables us to write for $N\to\infty$:
\small
\begin{eqnarray}
  \overline{\bra\sigma_{i}\sigma_{j}\ket^{\rho}}&=&
  \lim_{n\to 0}\lim_{N\to\infty}
     \frac{
     \sum_{\rho^\prime \rho^\pprime=0}^n
     \left[\frac{\lambda_{\rho^\prime}(n)}{\lambda_0(n)}\right]^{j-i}   \left[\frac{\lambda_{\rho^\pprime}(n)}{\lambda_0(n)}\right]^{N-j+i}
     \tr(\bS_{\{\rho\}}\bU_n^{(\rho^\prime)}\bS_{\{\rho\}}\bU_n^{(\rho^\pprime)})}
{1+ \sum_{\rho^\prime=1}^n
\left[\frac{\lambda_{\rho^\prime}(n)}{\lambda_0(n)}\right]^N }
\nonumber
\\
&=&
  \lim_{n\to 0}
       \sum_{\rho^\prime =0}^n
     \lambda_{\rho^\prime}(0)^{j-i}
     \tr(\bS_{\{\rho\}}\bU_n^{(\rho^\prime)}\bS_{\{\rho\}}\bU_n^{(0)})
\label{eq:cor_intermediate}
\end{eqnarray}
\normalsize
To work out the trace in (\ref{eq:cor_intermediate}) we write the
entries of our projection matrices as follows:
\be
U_n^{(\rho^\prime)}(\bsigma,\bsigma^\prime)=\sum_{\scriptsize\!\!\!\begin{array}{c}\{\varsigma\},\{\varsigma^\prime\}\\
|\{\varsigma\}|=|\{\varsigma^\prime\}|=\rho^\prime
\end{array}\!\!\!}
V^{(\rho^\prime)}_{\{\varsigma\},\{\varsigma^\prime\}}~
u_{\{\varsigma\}}(\bsigma)
v_{\{\varsigma^\prime\}}(\bsigma^\prime) \label{eq:Uexplicit} \ee
We may now write
\small
\begin{eqnarray}
  \tr(\bS_{\{\rho\}}\bU_n^{(\rho^\prime)}\bS_{\{\rho\}}\bU_n^{(0)})&=&
     \sum_{\bsigma, \bsigma^\prime}
     \Big[v_{\{0\}}(\bsigma)\!\prod_{\alpha\in\{\rho\}}\!\sigma_\alpha\Big]~U_n^{(\rho^\prime)}(\bsigma,\bsigma^\prime)~
   \Big[u_{\{0\}}(\bsigma^\prime)\!\prod_{\alpha\in\{\rho\}}\!\sigma^\prime_\alpha\Big]
   \nonumber
   \\
   &=&\sum_{\scriptsize\!\!\!\begin{array}{c}\{\varsigma\},\{\varsigma^\prime\}\\
|\{\varsigma\}|=|\{\varsigma^\prime\}|=\rho^\prime
\end{array}\!\!\!}
V^{(\rho^\prime)}_{\{\varsigma\},\{\varsigma^\prime\}}
A_{\{\rho\}}^{(\{0\},\{\varsigma\})}
A_{\{\rho\}}^{(\{\varsigma^\prime\},\{0\})}
\end{eqnarray}
\normalsize
with
\begin{eqnarray}
A_{\{\rho\}}^{(\{0\},\{\varsigma\})}&=& \sum_{\bsigma}
     v_{\{0\}}(\bsigma)u_{\{\varsigma\}}(\bsigma)\prod_{\alpha\in\{\rho\}}\sigma_\alpha
     \label{eq:A1}
     \\
A_{\{\rho\}}^{(\{\varsigma\},\{0\})}&=&
   \sum_{\bsigma} u_{\{0\}}(\bsigma) v_{\{\varsigma\}}(\bsigma)\prod_{\alpha\in\{\rho\}}
   \sigma_\alpha
   \label{eq:A2}
\end{eqnarray}
Our correlations (\ref{eq:cor_intermediate}) can apparently be
written in the simplified form
\begin{eqnarray}
  \overline{\bra\sigma_{i}\sigma_{j}\ket^{\rho}}&=&
  \lim_{n\to 0}~
       \sum_{\rho^\prime =0}^n
     \lambda_{\rho^\prime}(0)^{j-i}\!\!\!\!
\sum_{\scriptsize\!\!\!\begin{array}{c}\{\varsigma\},\{\varsigma^\prime\}\\
|\{\varsigma\}|=|\{\varsigma^\prime\}|=\rho^\prime
\end{array}\!\!\!}\!
V^{(\rho^\prime)}_{\{\varsigma\},\{\varsigma^\prime\}}
A_{\{\rho\}}^{(\{0\},\{\varsigma\})}
A_{\{\rho\}}^{(\{\varsigma^\prime\},\{0\})}
      \label{eq:correlations1}
\end{eqnarray}
 Inserting the eigenvectors (\ref{eq: right},\ref{eq: left})
into expressions (\ref{eq:A1},\ref{eq:A2}) for the coefficients
$A^{(\{0\},\{\varsigma\})}_{\{\rho\}}(n)$ and
$A^{(\{\varsigma\},\{0\})}_{\{\rho\}}(n)$, followed by summation
summing over the spin variables,  gives
\small
\begin{eqnarray}
  A^{(\{0\},\{\varsigma\})}_{\{\rho\}}(n)&=&
    \int\!\mathrm{d}x\mathrm{d}\mu~P_{\rho^\prime}(x,\mu\vert n)
          \int\!\mathrm{d}y~\Psi_{0}(y\vert n) \nonumber \\
  &&\quad\times
          [2\cosh(\beta x+\beta y)]^{n}
          [1-\mu\tanh(\beta x+\beta y)]^{\vert\{\rho\}\cap\{\varsigma\}\vert} \nonumber \\
  &&\quad\times
     [\tanh(\beta x+\beta y)]^{\vert\{\rho\}\cap\overline{\{\varsigma\}}\vert}
     [\tanh(\beta x+\beta y)-\mu]^{\vert\overline{\{\rho\}}\cap\{\varsigma\}\vert}
     \label{eq:BigA1}
\\
  A^{(\{\varsigma\},\{0\})}_{\{\rho\}}(n)&=&
  \int\!\mathrm{d}x~\Phi_{0}(x\vert n)
  \int\!\mathrm{d}y\mathrm{d}\nu~Q_{\rho^\prime}(y,\nu\vert n) \nonumber \\
  &&\quad\times
          [2\cosh(\beta x+\beta y)]^{n}
          [1-\nu\tanh(\beta x+\beta y)]^{\vert\{\rho\}\cap\{\varsigma\}\vert}\nonumber \\
  &&\quad\times
     [\tanh(\beta x+\beta y)]^{\vert\{\rho\}\cap\overline{\{\varsigma\}}\vert}
     [\tanh(\beta x+\beta y)-\nu]^{\vert\overline{\{\rho\}}\cap\{\varsigma\}\vert}
     \label{eq:BigA2}
\end{eqnarray}
\normalsize
These quantities no longer depend on the detailed realizations of
the index sets, but only on the sizes of these sets and of their
intersections.
 Let us denote the number of elements in the intersection of $\{\rho\}$ and $\{\varsigma\}$ by
$k=\vert\{\rho\}\cap\{\varsigma\}\vert$,
$k=0,\ldots,\textrm{min}\{\rho,\rho^\prime\}$ (since
$|\{\varsigma\}|=\rho^\prime$):
 \be
  \vert\{\rho\}\cap\{\varsigma\}\vert=k, \quad
  \vert\{\rho\}\cap\overline{\{\varsigma\}}\vert=\rho-k, \quad
  \vert\overline{\{\rho\}}\cap\{\varsigma\}\vert=\rho^\prime-k
  \label{eq:set_bookkeeping}
\ee
 with similar definitions in the case of
$\{\varsigma^\prime\}$, defining the variable $k^\prime$.
 We may now write (\ref{eq:correlations1}) as
\begin{eqnarray}
  \overline{\bra\sigma_{i}\sigma_{j}\ket^{\rho}}&=&
  \lim_{n\to 0}
       \sum_{\rho^\prime =0}^n
     \lambda_{\rho^\prime}(0)^{j-i}
     \sum_{k,k^\prime=0}^{{\rm min}\{\rho,\rho^\prime\}}
A_{\rho,k}^{(0,\rho^\prime)} A_{\rho,k^\prime}^{(\rho^\prime,0)}
\nonumber
\\
&&\times
\sum_{\scriptsize\!\begin{array}{c}\{\varsigma\},~|\{\varsigma\}|=\rho^\prime\\
\vert\{\rho\}\cap\{\varsigma\}\vert=k
\end{array}\!}
\sum_{\scriptsize\!\begin{array}{c}\{\varsigma^\prime\},~|\{\varsigma^\prime\}|=\rho^\prime\\
\vert\{\rho\}\cap\{\varsigma^\prime\}\vert=k^\prime
\end{array}\!}
V^{(\rho^\prime)}_{\{\varsigma\},\{\varsigma^\prime\}}
      \label{eq:correlations1b}
\end{eqnarray}
in which $A_{\rho,k}^{(0,\rho^\prime)}$ and
$A_{\rho,k}^{(\rho^\prime,0)}$ denote the $n\to 0$ limits of
(\ref{eq:BigA1}) and (\ref{eq:BigA2}), respectively (with the
conventions as laid down in (\ref{eq:set_bookkeeping})):
\begin{eqnarray}
  A^{(0,\rho^\prime)}_{\rho,k}&=&
    \int\!\mathrm{d}x\mathrm{d}\mu~P_{\rho^\prime}(x,\mu\vert 0)
          \int\!\mathrm{d}y~\Psi_{0}(y\vert 0)~  [\tanh(\beta x+\beta y)]^{\rho-k}\nonumber \\
  &&\times
          [1-\mu\tanh(\beta x+\beta y)]^{k}
     [\tanh(\beta x+\beta y)-\mu]^{\rho^\prime-k}
     \label{eq:BigA1n0}
\\
  A^{(\rho^\prime,0)}_{\rho,k}&=&
  \int\!\mathrm{d}x~\Phi_{0}(x\vert 0)
  \int\!\mathrm{d}y\mathrm{d}\nu~Q_{\rho^\prime}(y,\nu\vert 0)~  [\tanh(\beta x+\beta y)]^{\rho-k}\nonumber \\
  &&\times
          [1-\nu\tanh(\beta x+\beta y)]^{k}
     [\tanh(\beta x+\beta y)-\nu]^{\rho^\prime-k}
     \label{eq:BigA2n0}
\end{eqnarray}
The rigorous evaluation of the last line in
(\ref{eq:correlations1b}) for arbitrary models requires the
explicit calculation of the expansion factors
$V^{(\rho^\prime)}_{\{\varsigma\},\{\varsigma^\prime\}}$. Although
one can easily write formal expressions for these quantities in
terms of the inverse of the matrix of inner products of the
eigenvectors within a given eigenspace $\rho^\prime$, this leads
as yet only to expressions in which it is not clear how the limit
$n\to 0$ can be taken.

We can at present only push the evaluation of
(\ref{eq:correlations1b}) to its conclusion for those cases where
the eigenvectors within each eigenspace are either explicitly
orthogonal for any $n$ (as in chains without disorder, or in the
random bond chain without external fields), or become effectively
orthogonal in the $n\to 0$ limit. The latter is very hard to
verify or disprove {\em a priori}, but can serve as an efficient
ansatz, to be verified later using numerical simulations. In these
cases we are allowed to write simply
$V^{(\rho^\prime)}_{\{\varsigma\},\{\varsigma^\prime\}}=D_{\rho^\prime}^{-1}(n)\delta_{\{\varsigma\},\{\varsigma^\prime\}}$
and find (\ref{eq:correlations1b})  reducing to
\begin{eqnarray}
  \overline{\bra\sigma_{i}\sigma_{j}\ket^{\rho}}&=&\lim_{n\to 0}
   \bigg\{\sum_{\varsigma=0}^{\rho}\frac{\lambda_{\varsigma}(0)^{j-i}}{D_{\varsigma}(0)}
     \sum_{k=0}^{\varsigma}{\rho \choose k}{n-\rho \choose \varsigma-k}
      A_{\rho,k}^{(0,\varsigma)}A_{\rho,k}^{(\varsigma,0)}
  \nonumber \\
&&\hspace*{10mm}
  +\sum_{\varsigma>\rho}\frac{\lambda_{\varsigma}(0)^{j-i}}{D_{\varsigma}(0)}
     \sum_{k=0}^{\rho}{\rho \choose k}{n-\rho \choose \varsigma-k}
      A_{\rho,k}^{(0,\varsigma)}A_{\rho,k}^{(\varsigma,0)}\bigg\}
      \label{eq:correlations2}
\end{eqnarray}
 It turns out that in (\ref{eq:correlations2})
only the terms with $k=\varsigma$ will survive the limit $n\to 0$.
In the special case of non-disordered models, where
$P_{\rho}(x,\mu\vert n)=\delta(x-x^{*})\delta(\mu-\mu^{*})$ and
 $Q_{\rho}(y,\nu\vert n)=\delta(y-y^{*})\delta(\nu-\mu^{*})$, with $\mu^{*}=\tanh(\beta(x^{*}+y^{*}))$,
 we see that $A_{\rho,k}^{(0,\varsigma)}$ and
$A_{\rho,k}^{(\varsigma,0)}$ vanish unless $k=\varsigma$. More
generally we show in the Appendix that for integer $\rho$ and
$\ell$:
\be
\rho\geq 1,~\ell\geq 0:~~~~~~\lim_{n\to 0} {n-\rho \choose
\ell}=\delta_{\ell,0} \label{eq:combinatorics} \ee
 It follows that
the second line of (\ref{eq:correlations2}) must vanish entirely
since there always $k\leq \rho<\varsigma$, whereas in the first
line we retain only the terms with $k=\varsigma$, so that together
with  (\ref{eq:expressionforD}):
\small
\begin{eqnarray}
\hspace*{-18mm} \label{eq:corlapp}
  \overline{\bra\sigma_{i}\sigma_{j}\ket^{\rho}}&=&
  \sum_{\varsigma=0}^{\rho}D_{\varsigma}^{-1}{\rho \choose \varsigma}
    A^{(0,\varsigma)}_{\rho}A^{(\varsigma,0)}_{\rho}\lambda_{\varsigma}(0)^{j-i}
\\
\hspace*{-18mm}
  A^{(0,\varsigma)}_{\rho}&=&
    \int\!\mathrm{d}x\mathrm{d}\mu~P_{\varsigma}(x,\mu\vert 0)
          \int\!\mathrm{d}y~\Psi_{0}(y\vert 0)
          [1-\mu\tanh(\beta x+\beta y)]^{\varsigma}
   [\tanh(\beta x+\beta y)]^{\rho-\varsigma}
\\
\hspace*{-18mm}
  A^{(\varsigma,0)}_{\rho}&=&
  \int\!\mathrm{d}x~\Phi_{0}(x\vert 0)
  \int\!\mathrm{d}y\mathrm{d}\nu~Q_{\varsigma}(y,\nu\vert 0)
          [1-\nu\tanh(\beta x+\beta y)]^{\varsigma}
     [\tanh(\beta x+\beta y)]^{\rho-\varsigma}
\\
\hspace*{-18mm}
  D_{\rho}&=&
  \int\!\mathrm{d}x\mathrm{d}\mu~P_{\rho}(x,\mu\vert 0)
                \int\!\mathrm{d}y\mathrm{d}\nu~Q_{\rho}(y,\nu\vert 0)
    \Big[1+\mu\nu-\tanh(\beta x+\beta y)[\mu+\nu]\Big]^{\rho}
\end{eqnarray}
\normalsize
This concludes our calculations for the random field Ising chain.
\vsp

\begin{figure}[t]
 \setlength{\unitlength}{0.50mm}\hspace*{33mm}
\begin{picture}(200,300)
\put(70,290){\here{$m,q$}} \put(190,290){\here{$a_1,a_2,r$}}
\put(-25,241){\sl weak} \put(-25,230){\sl fields}

\put(-25,146){\sl medium}\put(-25,135){\sl fields}

\put(-25,51){\sl strong}\put(-25,40){\sl fields}

\put(15,190){\includegraphics[height=90\unitlength,width=110\unitlength]{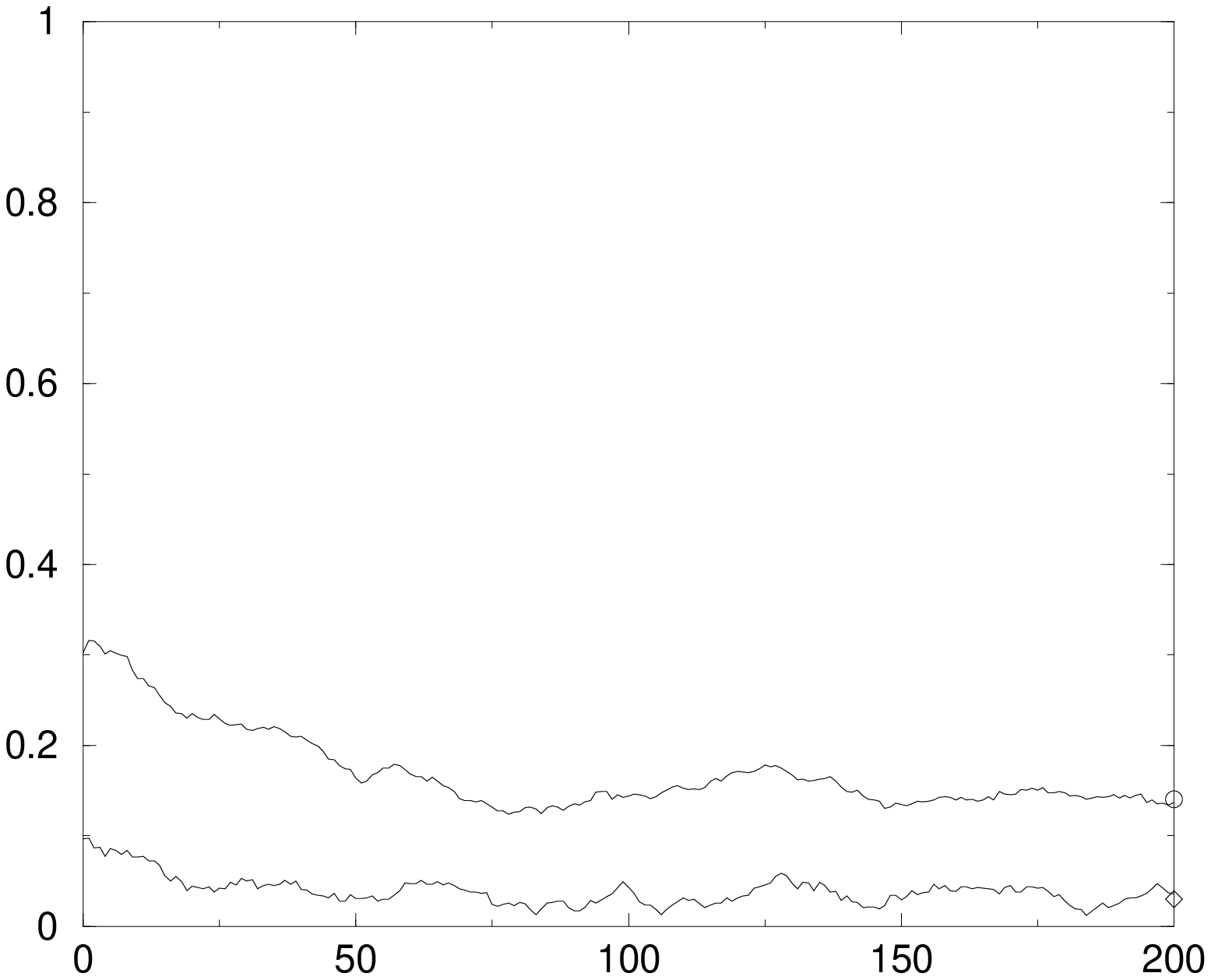}}
\put(135,190){\includegraphics[height=90\unitlength,width=110\unitlength]{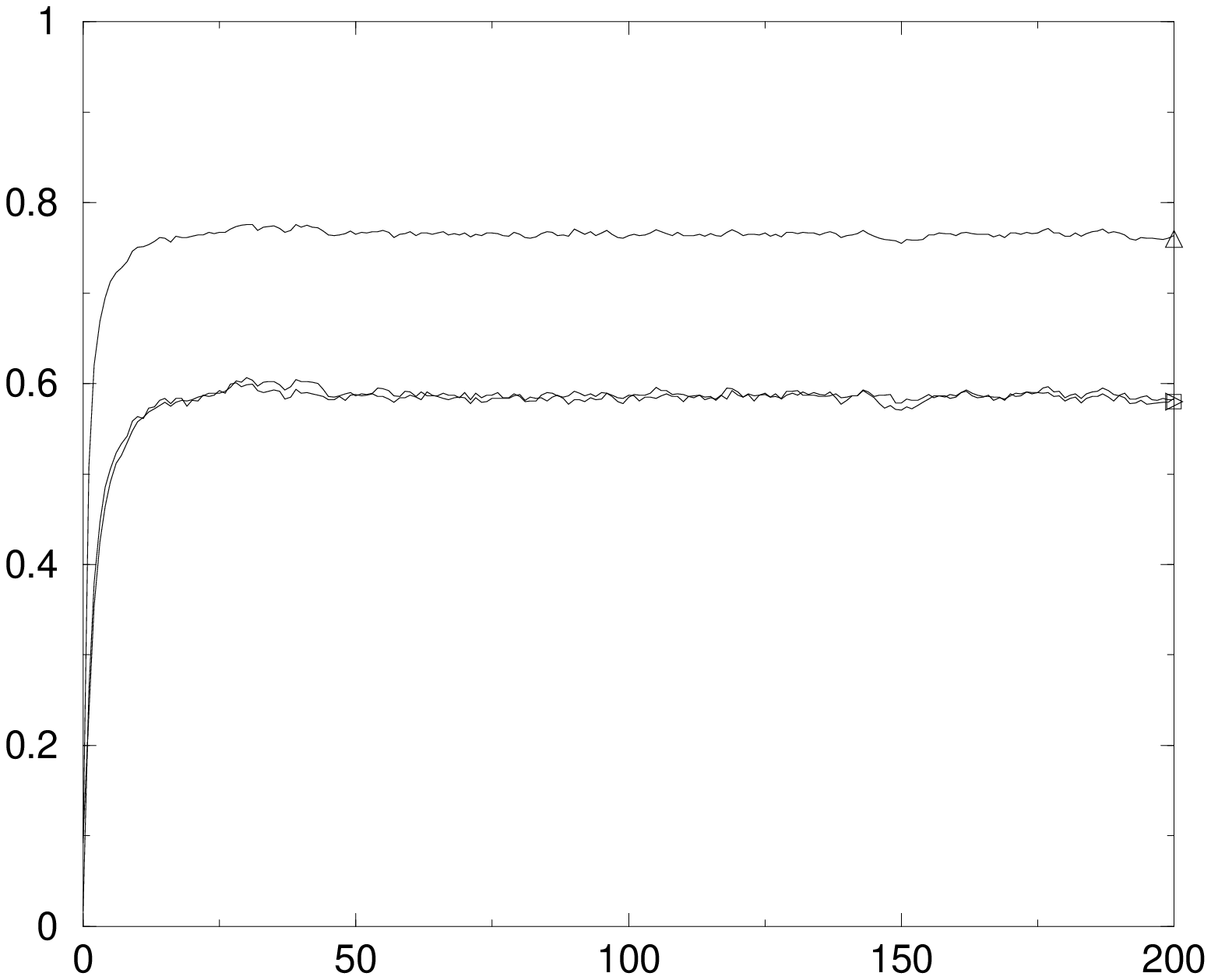}}
\put(15,95){\includegraphics[height=90\unitlength,width=110\unitlength]{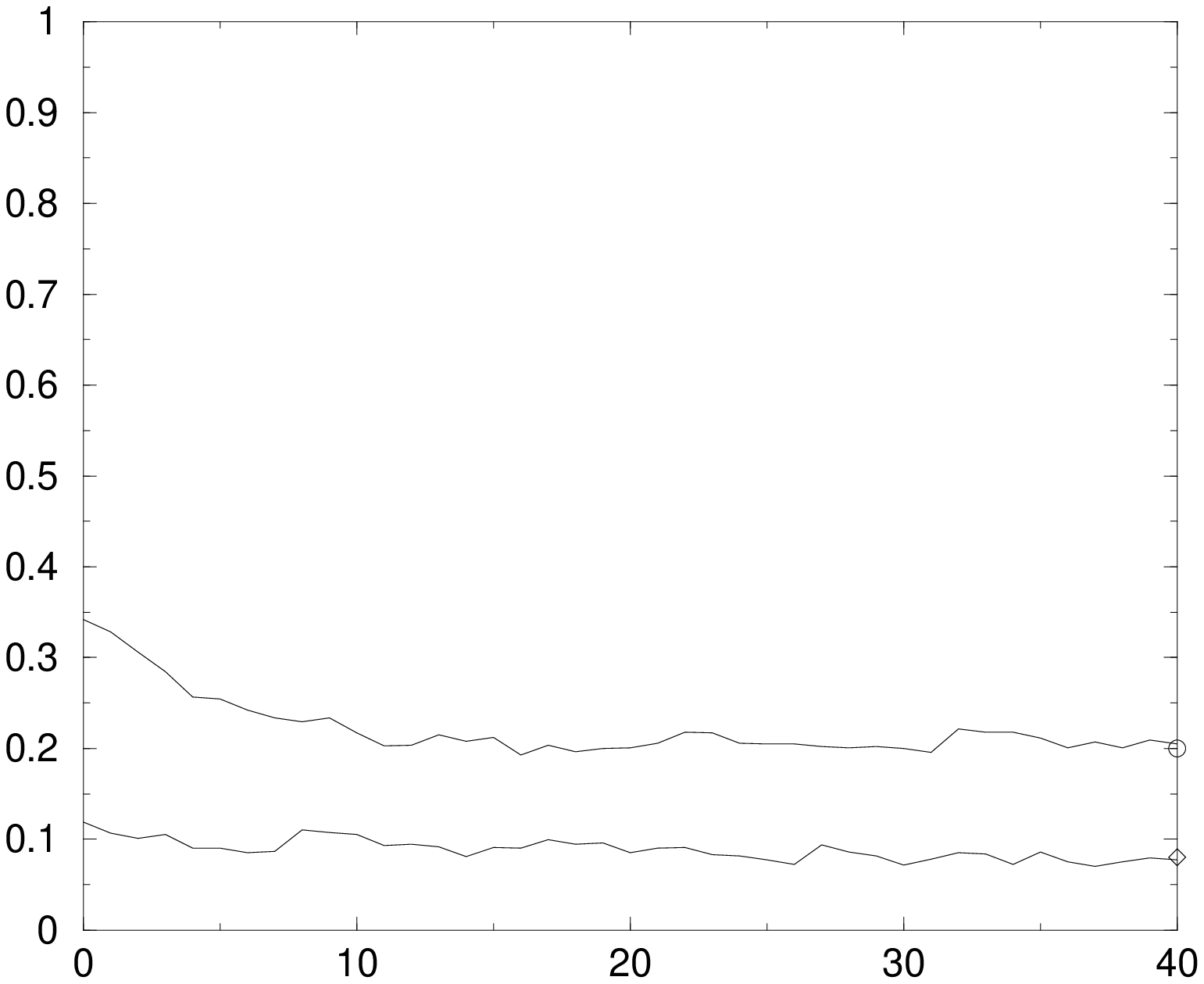}}
\put(135,95){\includegraphics[height=90\unitlength,width=110\unitlength]{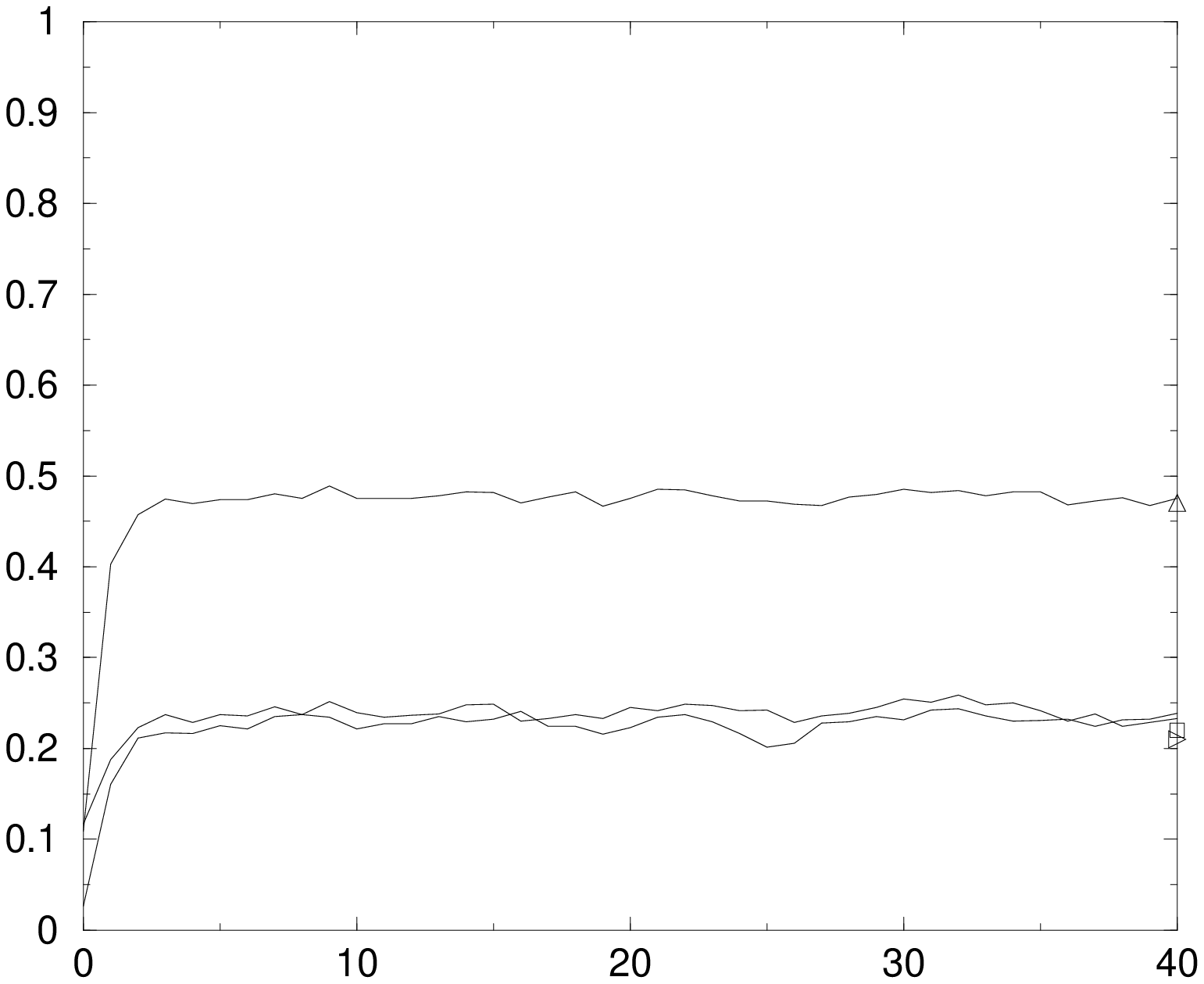}}
\put(12,0){\includegraphics[height=90\unitlength,width=113\unitlength]{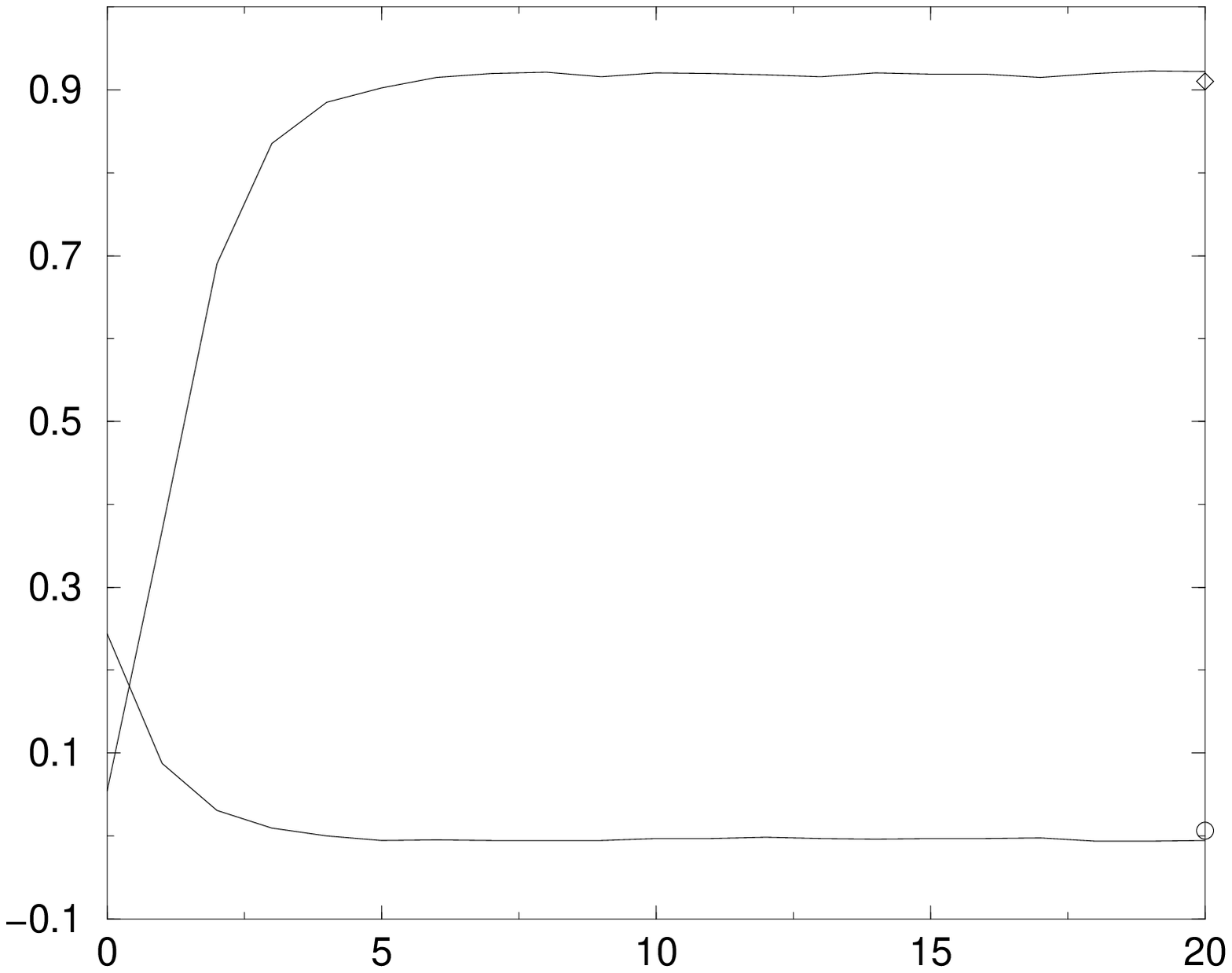}}
\put(132,0){\includegraphics[height=90\unitlength,width=113\unitlength]{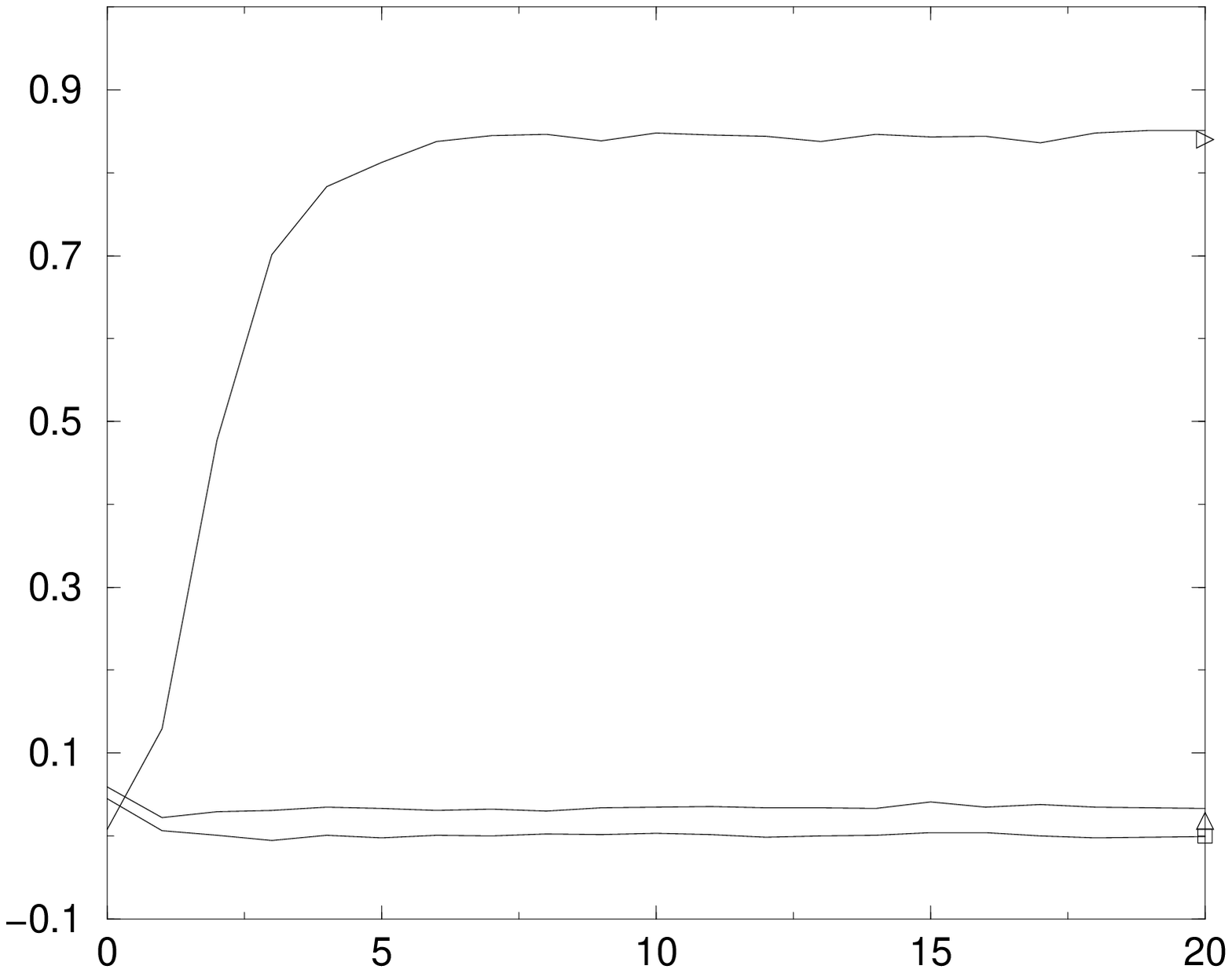}}
\put(70,-8){\here{\small $t$ (iter/spin)}}
\put(188,-8){\here{\small $t$ (iter/spin)}}
\end{picture}
\vspace*{8mm} \caption{Relaxation of observables towards
equilibrium at $T=1$, in two random field Ising chains with
identical disorder realizations, of size $N=20,\!000$ and with
field distribution
$p(\theta)=p\delta(\theta-\tilde{\theta})+(1-p)\delta(\theta+\tilde{\theta})$.
Left column: evolution of the magnetization
$m=N^{-1}\sum_i\sigma_i$ and the order parameter $q=N^{-1}\sum_i
\sigma_i\sigma_i^\prime$. Right column: evolution of the multiple
site quantities
  $a_{1}=N^{-1}\sum_{i}\sigma_{i}\sigma_{i+1}$,
  $a_{2}=N^{-1}\sum_{i}\sigma_{i}\sigma_{i+2}$, and
  $r=N^{-1}\sum_{i}\sigma_{i}\sigma_{i+1}\sigma_{i}^\prime\sigma_{i+1}^\prime$.
Different rows correspond to different control parameters.
  Top row: weak random fields, with $J_{0}=1$, $\tilde{\theta}=0.05$, $p=0.7$,  where the
         theoretical equilibrium predictions are $m\simeq 0.14$, $q\simeq 0.03$, $a_1\simeq 0.76$, $a_2\simeq 0.58$, $r\simeq 0.58$.
  Middle row: intermediate fields, with $J_{0}=0.5$, $\tilde{\theta}=0.2$, $p=0.7$,  where
  our theory predicts $m\simeq 0.20$, $q\simeq 0.08$, $a_1\simeq 0.47$, $a_2\simeq 0.22$, $r\simeq 0.21$.
  Bottom row: strong random fields, with
         $J_{0}=0.2$, $\tilde{\theta}=2$, $p=0.5$, where the theory
         predicts the equilibrium values $m\simeq 0.006$, $q\simeq 0.91$, $a_1\simeq 0.018$, $a_2\simeq 0.0003$, $r=0.84$.
In all cases the predictions are indicated by markers at the right
of the graphs.} \label{fig:sim_chain}
\end{figure}
\begin{figure}[t]
\setlength{\unitlength}{0.6mm}\hspace*{12mm}
\begin{picture}(200,260)
\put(0,164){\includegraphics[height=77\unitlength,width=80\unitlength]{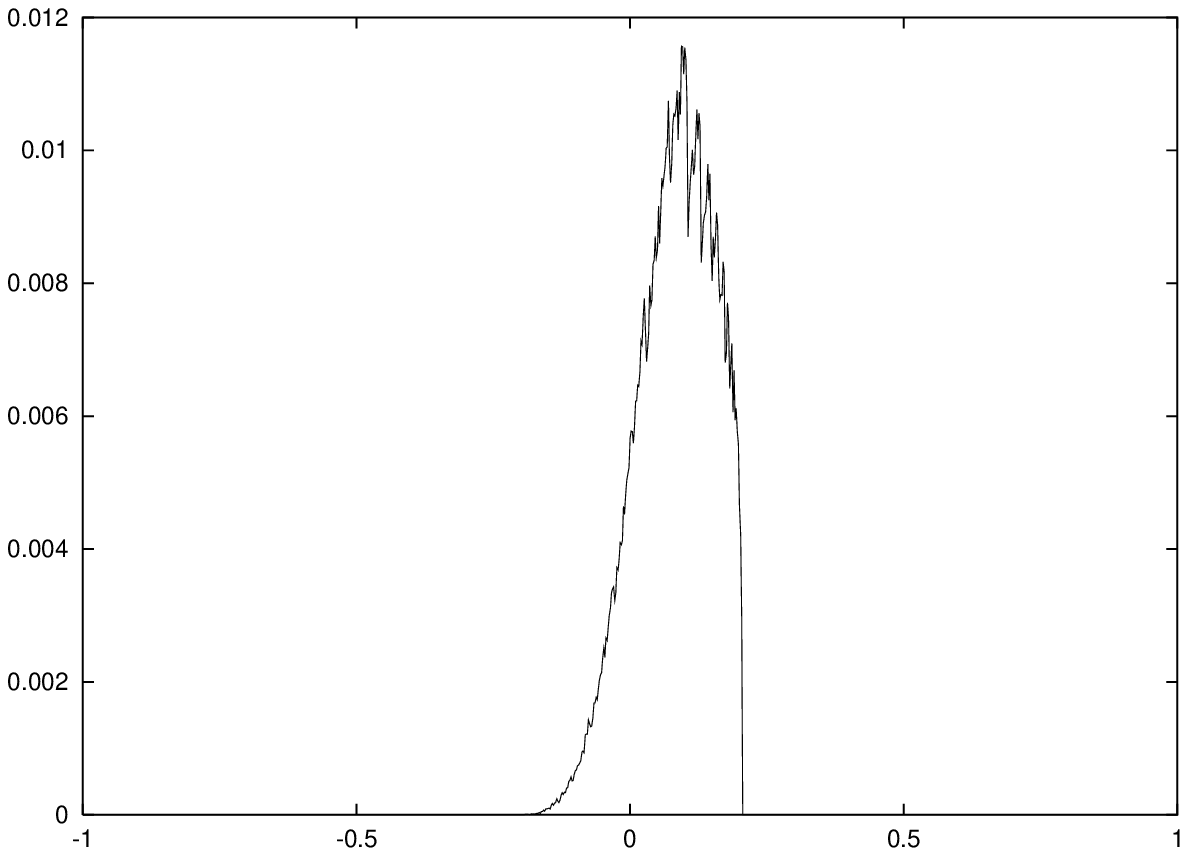}}
\put(80,164){\includegraphics[height=77\unitlength,width=80\unitlength]{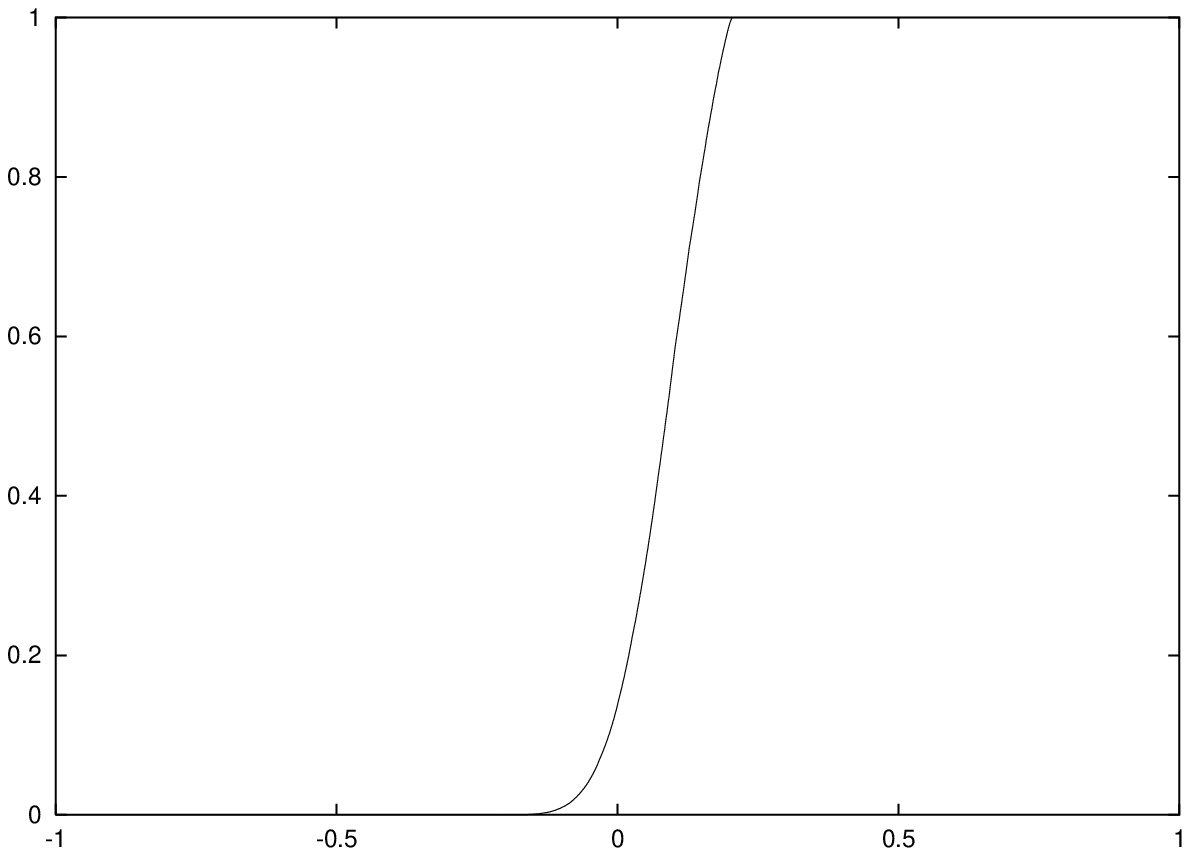}}
\put(160,164){\includegraphics[height=77\unitlength,width=80\unitlength]{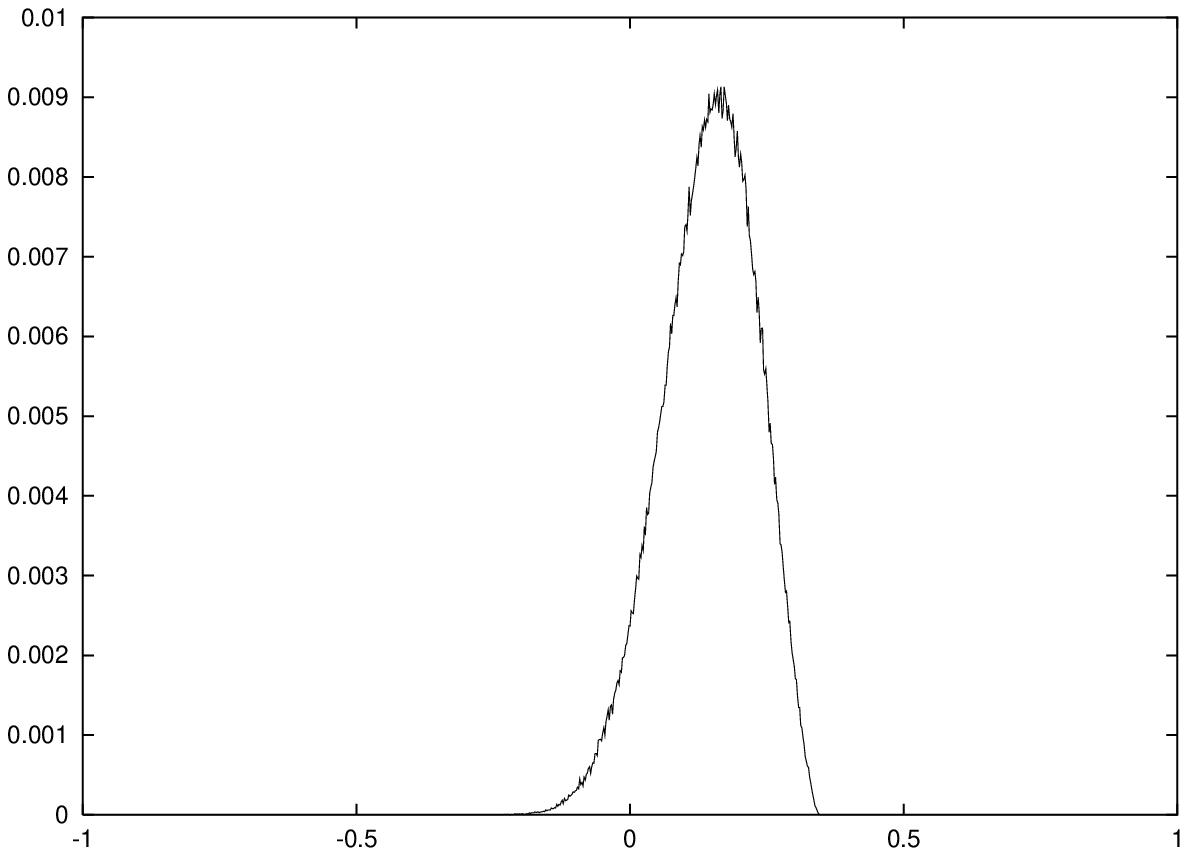}}
\put(0,82){\includegraphics[height=77\unitlength,width=80\unitlength]{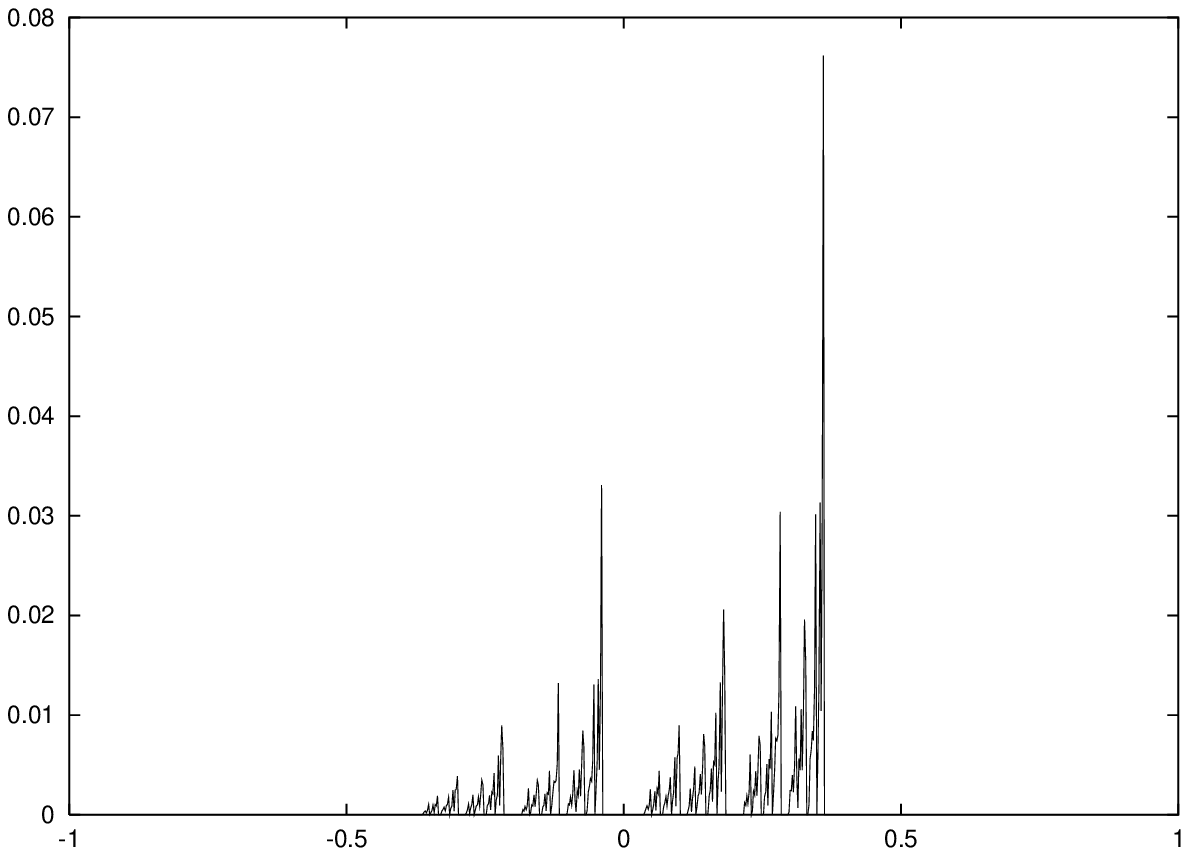}}
\put(80,82){\includegraphics[height=77\unitlength,width=80\unitlength]{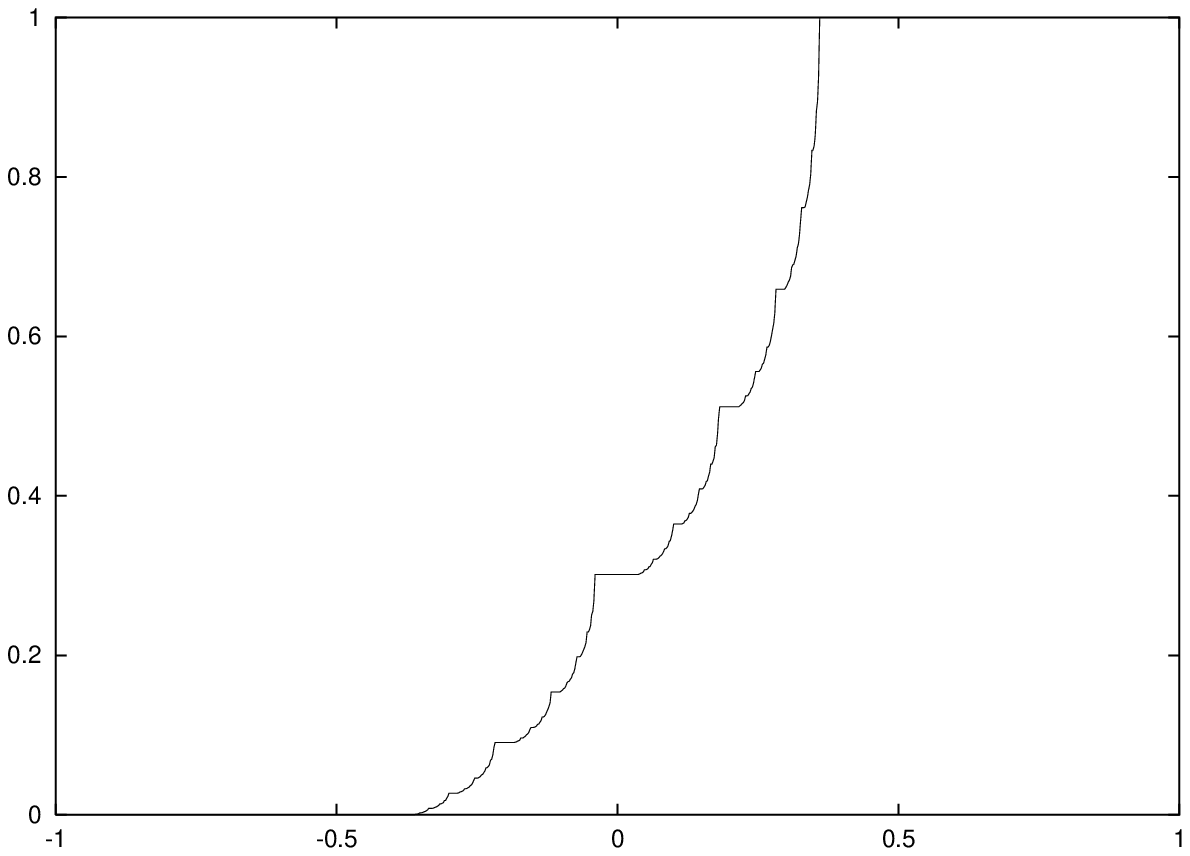}}
\put(160,82){\includegraphics[height=77\unitlength,width=80\unitlength]{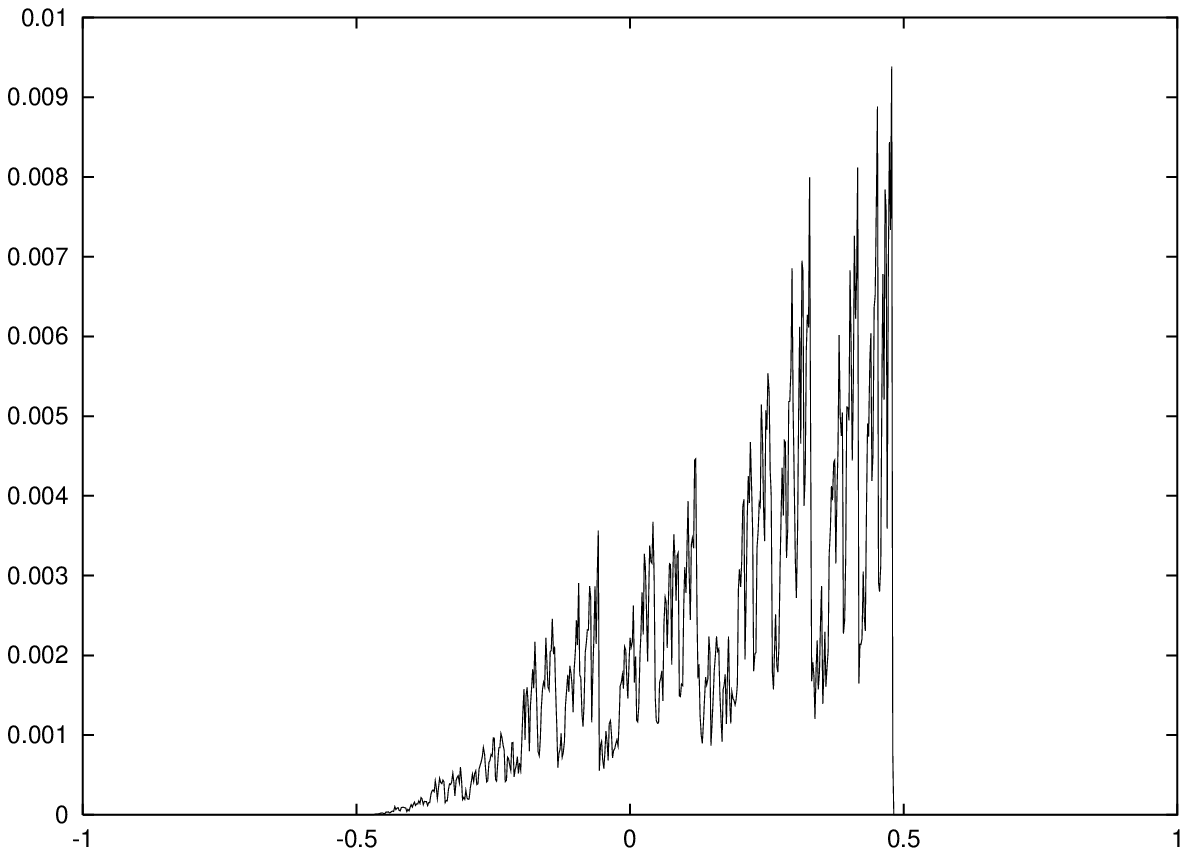}}
\put(0,0){\includegraphics[height=77\unitlength,width=80\unitlength]{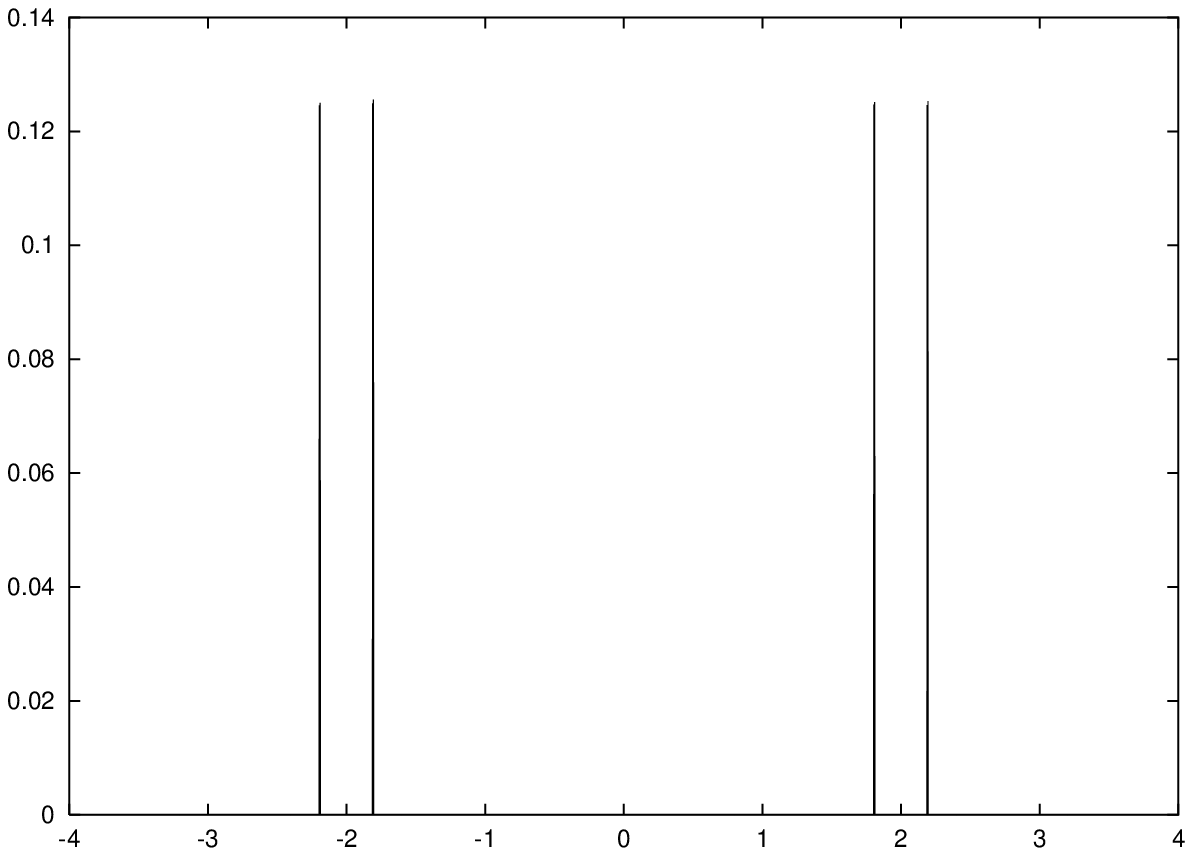}}
\put(80,0){\includegraphics[height=77\unitlength,width=80\unitlength]{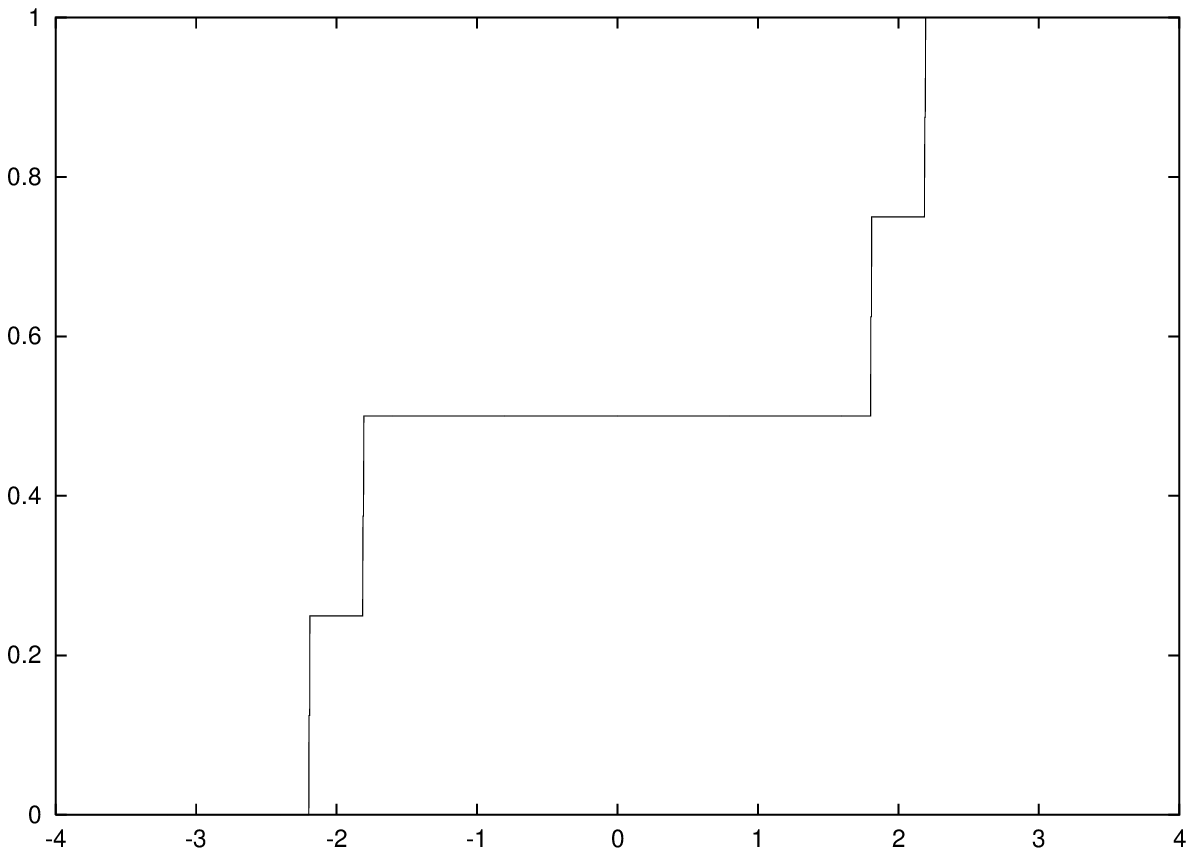}}
\put(160,0){\includegraphics[height=77\unitlength,width=80\unitlength]{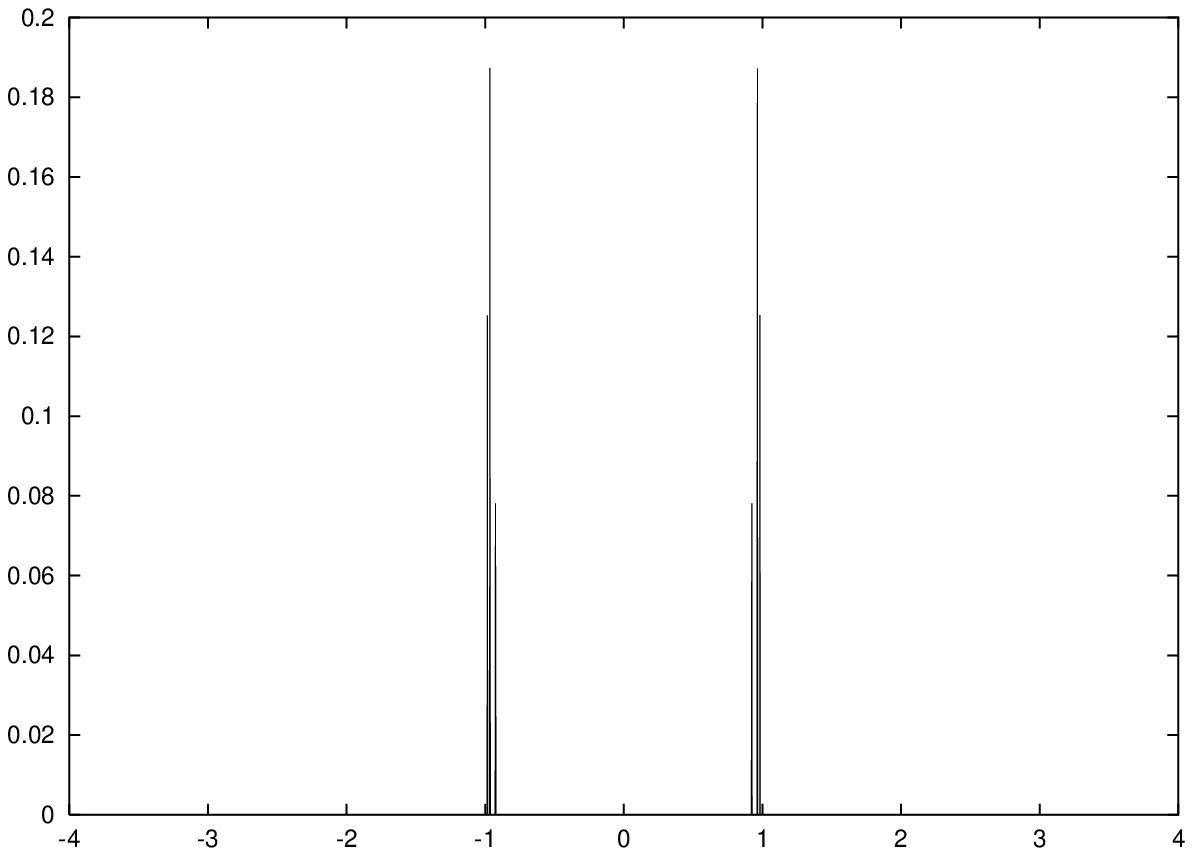}}
\put(42,-8){\here{\small $x$}} \put(122,-8){\here{\small$x$}}
\put(202,-8){\here{\small $m$}} \put(42,250){\here{$\Phi(x)$}}
\put(122,250){\here{$\hat{\Phi}(x)$}} \put(202,250){\here{$W(m)$}}
\end{picture}
\vspace*{8mm} \caption{Field distributions corresponding to the
data of the previous figure \ref{fig:sim_chain}, as obtained by
numerical solution of our integral eigenvalue equation (\ref{eq:
PHI}) via a population dynamics algorithm. The rows correspond to
again to weak random fields (top row), intermediate random fields
(middle row), and strong random fields (bottom row). Left column:
The effective field distribution $\Phi(x)$. Middle column:
         the integrated distribution
         $\hat{\Phi}(x)=\int_{-\infty}^{x}\!\mathrm{d}z~\Phi(z)$.
         Right column: the distribution of single-site magnetizations
         $W(m)=\int\!\mathrm{d}x\mathrm{d}y~\Phi(x)\Psi(y)\delta[m-\tanh(\beta x+\beta y)]$.}
         \label{fig:th_chain}
\end{figure}

\subsection{Comparison with simulations}

 We have tested the predictions  (\ref{eq:magn},\ref{eq:corlapp}) for the random field Ising chain with
 $p(\theta)=p\delta(\theta-\tilde{\theta})+(1-p)\delta(\theta+\tilde{\theta})$.
 Objects such as $\bra\sigma_i\ket^2$ or $\bra
 \sigma_i\sigma_j\ket^2$ were
 measured by simulating two copies of the system, with identical disorder realizations but each evolving independently
according to standard Glauber dynamics towards equilibrium
following a randomly chosen microscopic initial state. The results
are shown in figure \ref{fig:sim_chain}. In all simulations the
system size was $N=20,\!000$ spins. We concentrated on the
following quantities:
 \begin{eqnarray}
  m=\frac{1}{N}\sum_{i}\bra\sigma_{i}\ket,\quad
  a_{1}=\frac{1}{N}\sum_{i}\bra \sigma_{i}\sigma_{i+1}\ket,\quad
  a_{2}=\frac{1}{N}\sum_{i}\bra \sigma_{i}\sigma_{i+2}\ket
  \label{eq:single_sys}
\\
  q=\frac{1}{N}\sum_{i}\bra \sigma_{i}\ket^2, \quad
  r=\frac{1}{N}\sum_{i}\bra \sigma_{i}\sigma_{i+1}\ket^2
  \label{eq:double_sys}
\end{eqnarray}
The evaluation of the theoretical predictions
(\ref{eq:magn},\ref{eq:corlapp}) involved solving the relevant
functional eigenvalue equations numerically. For $m$ and $q$,
which both follow from (\ref{eq:magn}), one just needs to solve
(\ref{eq: PHI}) for $\lambda_0(0)=1$, which is straightforward
(either by iteration, or using a population dynamics algorithm).
The function $\Psi(y)$ subsequently follows via identity (\ref{eq:
PHI-PSI}).
 We see
in figure \ref{fig:sim_chain} that for $m$ and $q$ the agreement
between theory and experiment is excellent. Figure
\ref{fig:th_chain} shows the corresponding shapes of the
distribution $\Phi(x)$ as well as the associated integrated
distribution and the distribution $W(m)$ of single-site
magnetizations, which show the by now familiar characteristics of
random field Ising models (see e.g.
\cite{BrandtGross78,BruinsmaAeppli83,gyorgyi-rujan84,evangelou87}).

 For those
observables which require evaluation of (\ref{eq:corlapp}), and
therefore numerical solution of the eigenvalue problems (\ref{eq:
P},\ref{eq: Q}) for different values of $\rho$ (which is feasible
but extremely demanding in computing time), we have used the
approximation consisting of replacing $P_\rho(\ldots)$ and
$Q_{\rho}(\ldots)$ for $\rho>0$ by $P_0(\ldots)$ and
$Q_{0}(\ldots)$, respectively. This would formally be allowed only
in the non-disordered case (where also the assumed orthogonality
of our eigenvectors within eigenspaces is correct), but is seen to
give surprisingly accurate results even for those cases where the
shape of these distributions is highly non trivial; see figures
\ref{fig:sim_chain} and \ref{fig:th_chain}.

\section{Applications of the theory: neural networks and `small
world' systems}

 The theory in section \ref{sec:theory} can
be applied to any model which involves replicated transfer
matrices. Here we demonstrate  how it may be used to analyze
models which are structurally  different from the random field
Ising model, in having not only short-range but also long-range
bonds.

\subsection{$1+\infty$ dimensional attractor neural
networks}

We now turn to the attractor neural network described by the
Hamiltonian (\ref{eq:NNH}), where short range interactions compete
with long-range ones. A detailed study of the model, based on the
more conventional methods of \cite{BrandtGross78,BruinsmaAeppli83}
can be found in \cite{skantzos-coolen00,skantzos-coolen01}; here
our objective is only to demonstrate how the present replicated
transfer matrix diagonalization formalism can also be put to use
in the context of such models. Upon introducing the $p$ overlap
order parameters
$m_{\mu}(\bsigma)=N^{-1}\sum_{i}\xi_{i}^{\mu}\sigma_{i}$, each of
which which measure the similarity between  the system's
microscopic configuration $\bsigma$ and a given stored pattern,
one arrives  after some standard manipulations at the following
expression for the partition function
\be
  Z=\int\!\mathrm{d}\bm~
    e^{N[-\frac{1}{2}\beta J_{\ell}\bm^{2}+r(\bm)]}
\ee
 where $\bm=(m_{1},\ldots,m_{p})$, $\bm^{2}=\sum_\mu m_\mu^2$, and
$r(\bm)=\frac{1}{N}\log R(\bm)$ with \be
  R(\bm)=\sum_{\sigma_{1}\ldots\sigma_{N}}
   e^{\beta J_{s}\sum_{i}\sigma_{i}\sigma_{i+1}(\bxi_{i}\cdot\bxi_{i+1})+
      \beta J_{\ell}\sum_{i}\sigma_{i}(\bm\cdot\bxi_{i})}
\ee
 One may now calculate $r(\bm)$ by regarding the random patterns
as disorder
 and use the replica approach to calculate the disorder
average. In the thermodynamic limit,  $r(\bm)$ (which is itself
mathematically identical to the free energy per spin of a suitably
defined chain) must be identical to its disorder average, with
probability one. Therefore we consider \bd
  \overline{r(\bm)}=\lim_{n\to 0}\frac{1}{n}
     \lim_{N\to\infty}\frac{1}{N}\log\overline{R^{n}(\bm)}
\ed
In particular we have
\begin{eqnarray*}
  \overline{R^{n}(\bm)}&=&2^{-pN}\sum_{\bxi_{1}\ldots\bxi_{N}}\sum_{\bsigma_{1}\ldots\bsigma_{N}}
    \prod_{i}e^{\beta J_{s}(\bsigma_{i}\cdot\bsigma_{i+1})(\bxi_{i}\cdot\bxi_{i+1})+
                \beta J_{\ell}(\bm\cdot\bxi_{i})\sum_{\alpha=1}^{n}\sigma_{i}^{\alpha}}  \\
   &=&\tr(\bT^{N}(\bm))
\end{eqnarray*}
where $\bsigma_i=(\sigma_i^1,\ldots,\sigma_i^n)$, and $\bT(\bm)$
is a $2^{np}\times 2^{np}$ transfer matrix with entries
\be
  T_{\bxi,\bxi^\prime}(\bsigma,\bsigma^\prime;\bm)= 2^{-p}e^{\beta J_{s}(\bxi\cdot\bxi^\prime)(\bsigma\cdot\bsigma^\prime)+
      \beta J_{\ell}(\bm\cdot\bxi)\sum_{\alpha=1}^{n}\sigma_{\alpha}}
\ee In order to determine the largest eigenvalue of this
replicated transfer matrix we make the by now familiar type of
ansatz for the its left and right eigenvector:
\be
  v_{\bxi}(\bsigma)=\int\!\mathrm{d}y~\Psi_{\bxi}(y\vert n)e^{\beta y\sum_{\alpha=1}^{n}\sigma_{\alpha}}
  \quad
  u_{\bxi}(\bsigma)=\int\!\mathrm{d}x~\Phi_{\bxi}(x\vert n)e^{\beta x\sum_{\alpha=1}^{n}\sigma_{\alpha}}
  \label{eq:ev_ANN}
\ee
 Our motivation for this particular choice of the dependence on the pattern vectors $\bxi$ is that for $p=1$
the dependence on the remaining pattern can be transformed away by
the gauge transformation $\bsigma_i\to\xi_i\bsigma_i$. This would
leave  a replicated transfer matrix of an Ising chain with
constant bonds, where the role of the external field is played by
$J_{\ell}m$. Thus for $p=1$  the present eigenvectors must reduce
to those as studied in section \ref{sec:eigenvectors}. Secondly,
the group (\ref{eq:ev_ANN}) obviously represents only a subset of
all eigenvectors (to be precise: the $\rho=0$ family, in the
language of the previous section). Building the full set is
straightforward, but here we restrict ourselves for brevity to the
main ones, i.e. those which control the free energy and the
single-site observables (the others only play a role when
calculating multiple-site observables).

Having introduced our eigenvectors, we proceed as in the random
field Ising model, adding $\bm$ as a conditioning label wherever
needed. We then find in the limit $n\to 0$ that
$\lambda(0;\bm)=1$, that our final eigenvalue problems are
defined in terms of  joint field-pattern distributions:
\begin{eqnarray}
  \Psi_{\bxi}(y\vert 0)&=& 2^{-p} \sum_{\bxi^\prime}\int\!\mathrm{d}y^\prime~\Psi_{\bxi^\prime}(y^\prime\vert
  0)~
    \delta[y-A(J_{s}(\bxi\cdot\bxi^\prime),y^\prime+J_{\ell}(\bm\cdot\bxi^\prime))]
\\
  \Phi_{\bxi}(x\vert 0)&=& 2^{-p} \sum_{\bxi^\prime}\int\!\mathrm{d}x^\prime~\Phi_{\bxi^\prime}(x^\prime\vert
  0)~
   \delta[x-J_{\ell}(\bm\cdot\bxi)-A(J_{s}(\bxi\cdot\bxi^\prime),x^\prime)]
\end{eqnarray}
These distributions are normalized according to
$\int\!\mathrm{d}x~\Phi_{\bxi}(x|0)=\int\!\mathrm{d}y~\Psi_{\bxi}(y|0)=1$
for all $\bxi$.
 The actual value
to be inserted for the vector $\bm$ in the above expressions is to
be solved from the saddle-point equations which determine the
stationary point of the extensive exponent in the partition sum.
This equation can simply be written as
$\bm=\lim_{N\to\infty}N^{-1}\sum_i\overline{\bra\sigma_{i}\bxi_{i}^{\mu}\ket}$.
Upon repeating the steps taken earlier in solving the random field
Ising model, we get:
\begin{eqnarray}
\label{eq: saddle_point}
  m_{\mu}&=&\lim_{n\to 0}\frac{\tr(\bS^{\mu}_{\{1\}}\bT^{N}(\bm))}{\tr(\bT^{N}(\bm))}
  \nonumber \\
  &=&2^{-p}\sum_{\bxi}\int\mathrm{d}x\mathrm{d}y\,\Phi_{\bxi}(x\vert 0)\Psi_{\bxi}(y\vert 0)
   \xi_{\mu}\tanh(\beta x+\beta y)
\end{eqnarray}
in which $\bS_{\{1\}}^{\mu}$ is a diagonal $2^{np}\times 2^{np}$
matrix with elements: \bd
  S_{\{1\},\bxi\bxi'}^{\mu}(\bsigma,\bsigma')=\delta_{\bxi,\bxi^\prime}\delta_{\bsigma,\bsigma^\prime}
  \xi_{\mu} \sigma_{1}
\ed The $n=0$ eigenvalue problems for $\Phi_{\bxi}$ and
$\Psi_{\bxi}$ are coupled to the saddle point equations for the
`mean field' order parameters. This feature is typical, within the
replica formalism, for all models where a one-dimensional
structure is embedded in a mean-field (or range-free)
architecture, as is the case here.

In order to calculate the free energy we need to know the
$\order(n)$ contribution $\lambda(\bm)$ to $\lambda(n;\bm)$ (i.e.
$\lambda(n;\bm)=1+n\lambda(\bm)+\order(n^2)$). The latter can be
expressed in terms of the $n=0$ effective field distributions, and
is found to be given by: \bd
  \lambda(\bm)=2^{-2p}\sum_{\bxi,\bxi^\prime}\int\!\mathrm{d}y~\Psi_{\bxi^\prime}(y\vert 0)
    \beta B(J_{s}(\bxi\cdot\bxi^\prime),y+J_{\ell}(\bm\cdot\bxi'))
\ed
Hence
\begin{eqnarray}
  \overline{r(\bm)}&=&\lim_{n\to 0}\frac{1}{n}\lim_{N\to\infty}\frac{1}{N}\log\lambda^{N}(n;\bm)
 \nonumber \\
    &=&\lim_{n\to 0}\frac{1}{n}\log[1+n\lambda(\bm)+\order(n^{2})]
     =\lambda(\bm)
\end{eqnarray}
Substitution of this result for $r(\bm)$ into the partition leads
to to our final result
\be
  f=\frac{1}{2}J_{\ell}\bm^2 -T\lambda(\bm)
\ee in which $\bm$ is given by the solution of (\ref{eq:
saddle_point}). The link with the results of
\cite{skantzos-coolen00} is can now be established  upon defining
a new random variable $k$, which in \cite{skantzos-coolen00}
represents the ratio of conditioned partition functions, and is
subject to a random non-linear map as one builds up the chain
iteratively from $N=1$ to $N=\infty$. With the following
definition the two solutions (the one in \cite{skantzos-coolen00}
and the one in this paper) become fully identical: \be
  P(k,\bxi)=2^{-p}\int\mathrm{d}y\,\Psi_{\bxi}(y)\delta[k-e^{-2\beta y}]
\ee

\subsection{`Small-world' ferromagnets}

\begin{figure}[t]
\vspace*{7mm} \setlength{\unitlength}{0.5mm}\hspace*{33mm}
\begin{picture}(200,185)
\put(70,190){\here{$m,q$}} \put(190,190){\here{$a_1,a_2,r$}}

\put(-40,157){\sl high} \put(-40,146){\sl
Poissonnian}\put(-40,135){\sl connectivity}

\put(-40,62){\sl low} \put(-40,51){\sl
Poissonnian}\put(-40,40){\sl connectivity}

\put(15,95){\includegraphics[height=90\unitlength,width=110\unitlength]{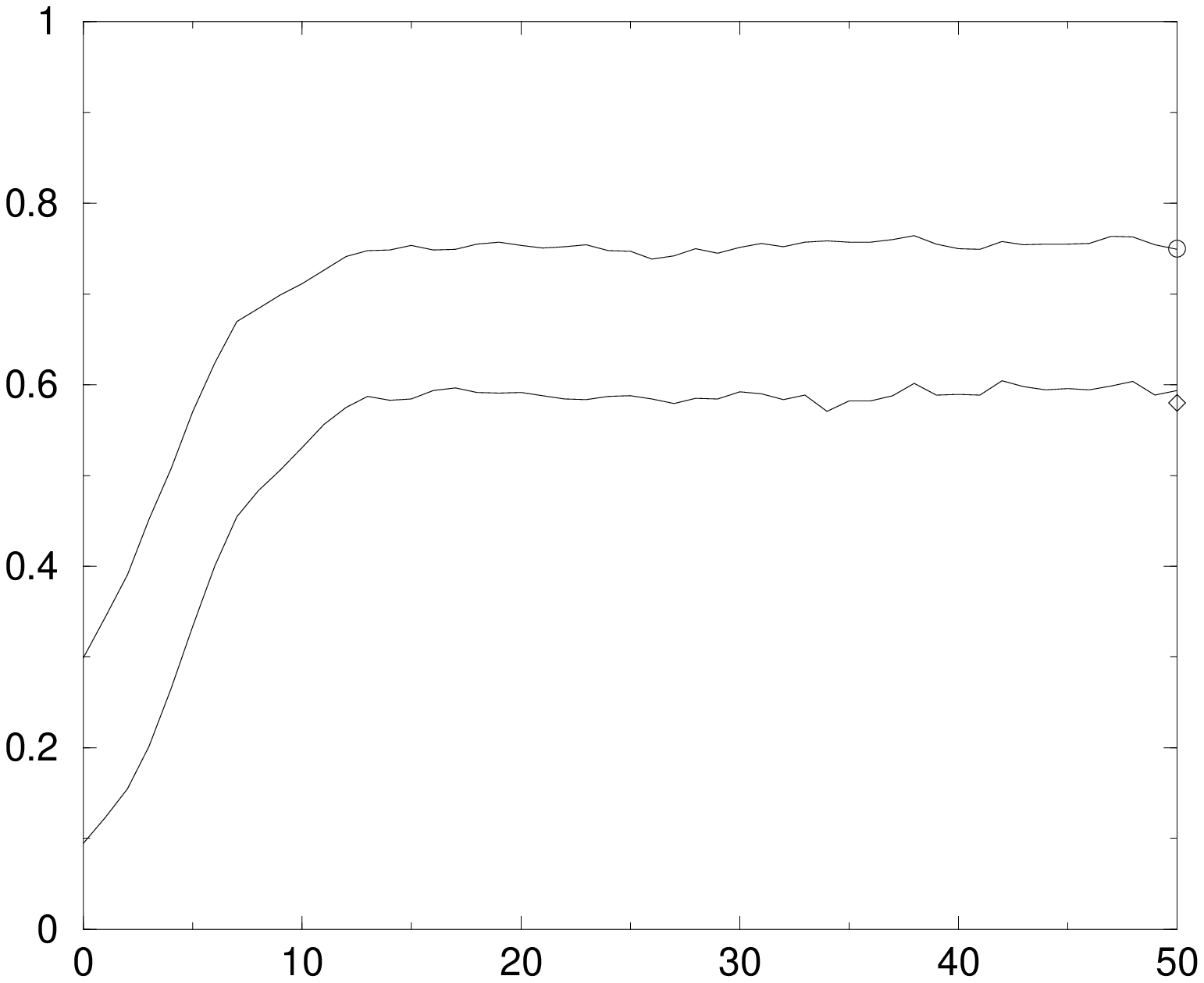}}
\put(135,95){\includegraphics[height=90\unitlength,width=110\unitlength]{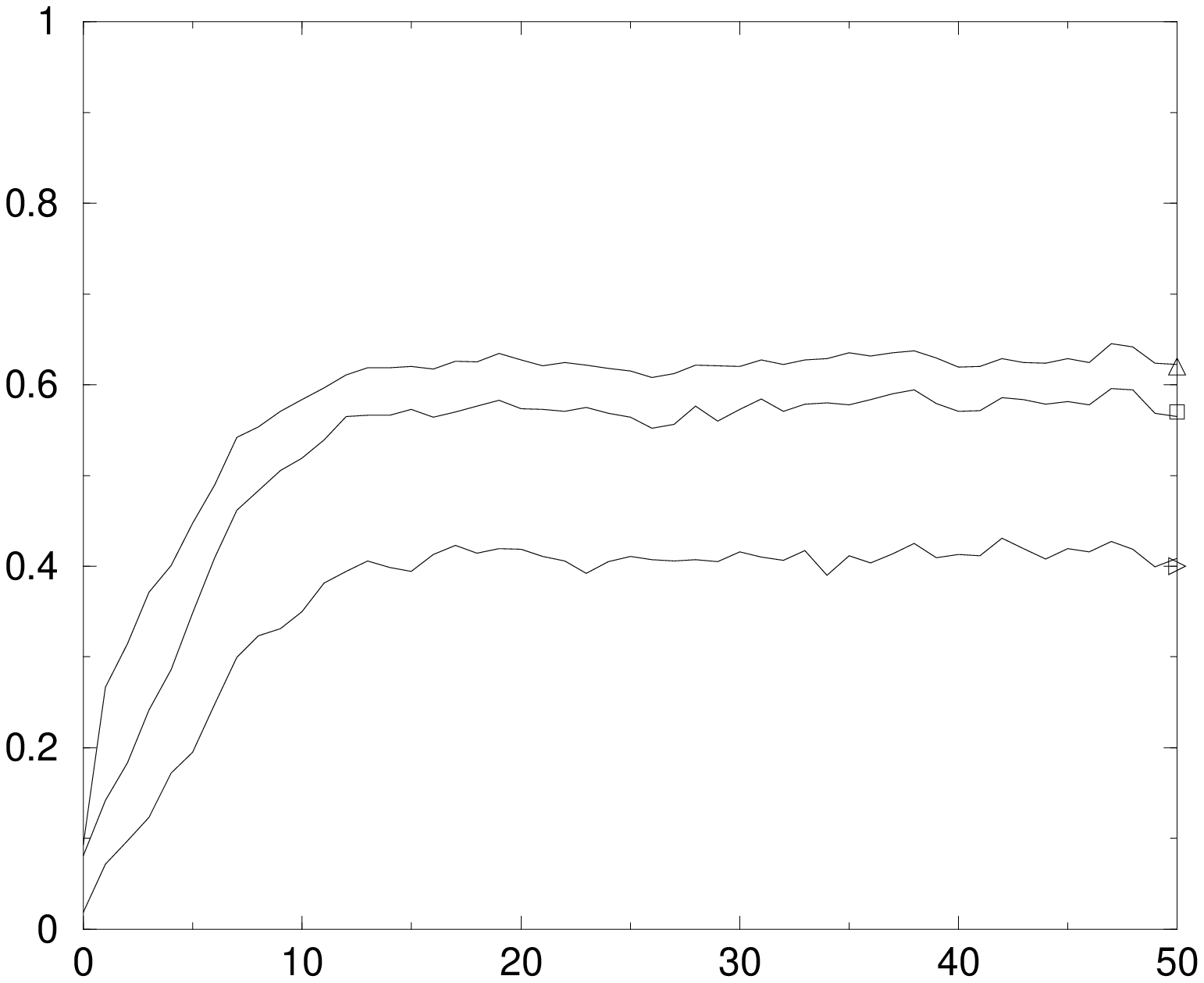}}
\put(15,0){\includegraphics[height=90\unitlength,width=110\unitlength]{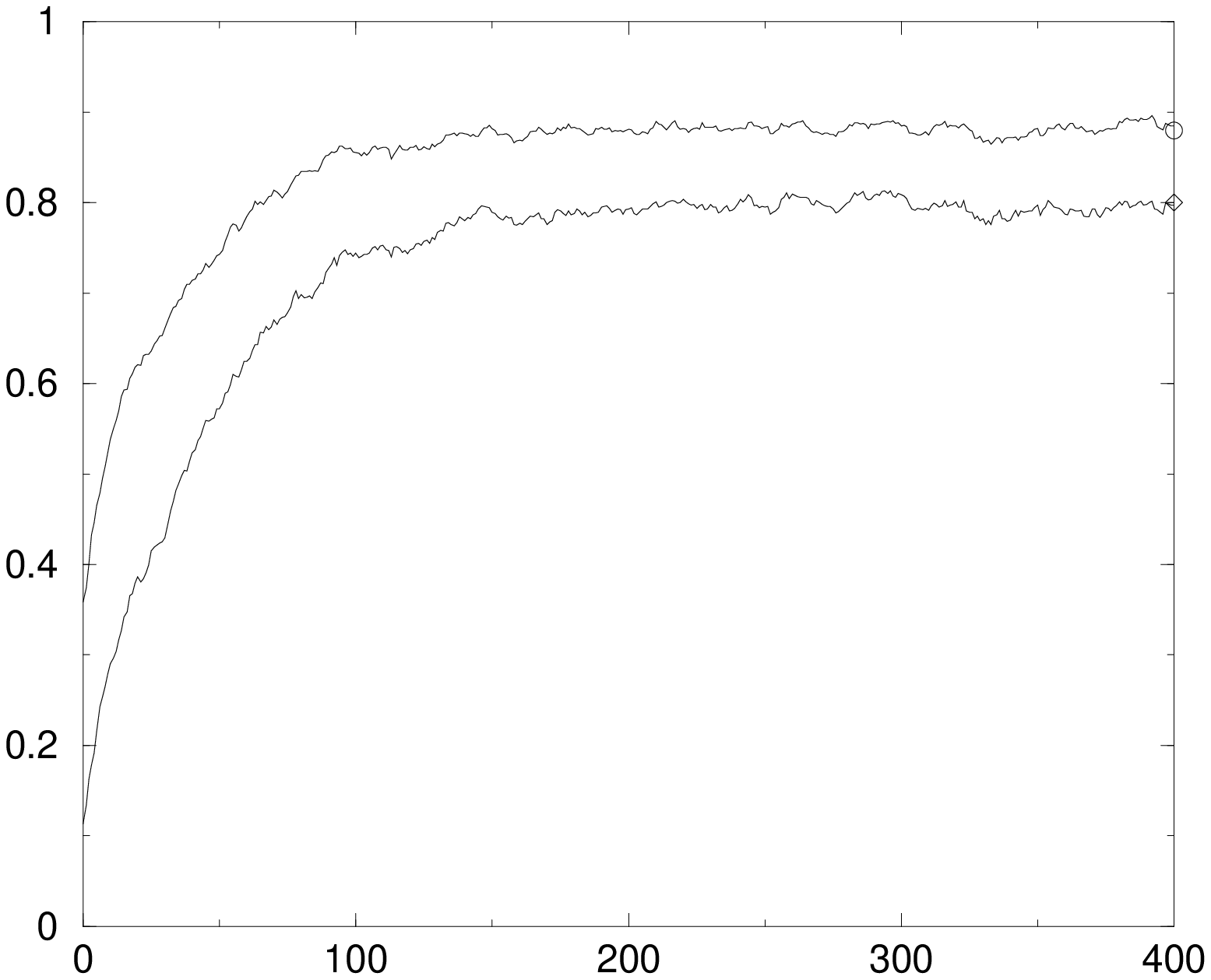}}
\put(135,0){\includegraphics[height=90\unitlength,width=110\unitlength]{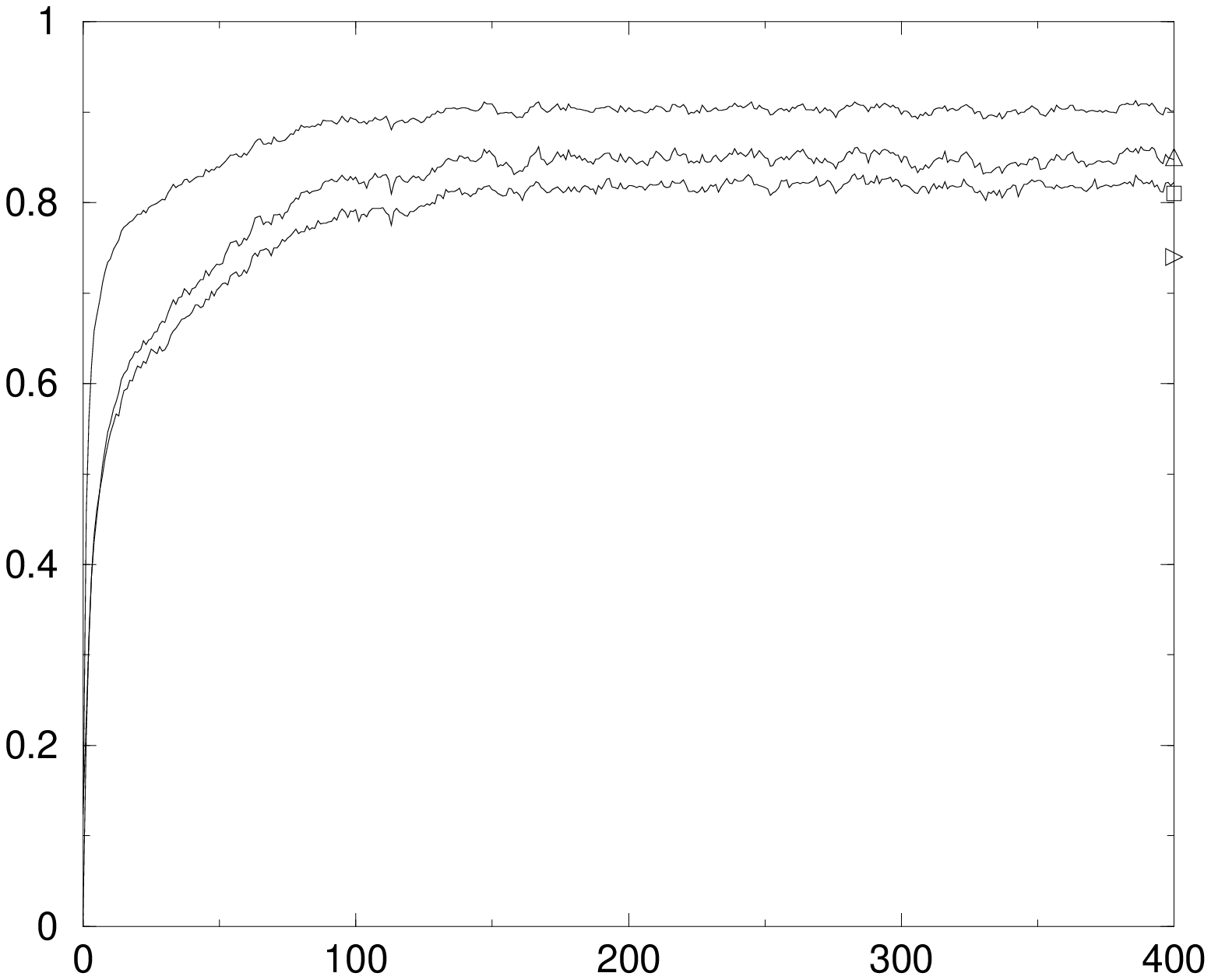}}
\put(70,-8){\here{\small $t$ (iter/spin)}}
\put(188,-8){\here{\small $t$ (iter/spin)}}
\end{picture}
\vspace*{10mm} \caption{ Relaxation of observables towards
equilibrium at $T=J=1$, in two `small world' ferromagnets with
identical realizations of the disorder (i.e. the Poissonnian
graph), of size $N=20,\!000$. Left column: evolution of the
magnetization $m=N^{-1}\sum_i\sigma_i$ and the order parameter
$q=N^{-1}\sum_i \sigma_i\sigma_i^\prime$. Right column: evolution
of the multiple site quantities
  $a_{1}=N^{-1}\sum_{i}\sigma_{i}\sigma_{i+1}$,
  $a_{2}=N^{-1}\sum_{i}\sigma_{i}\sigma_{i+2}$, and
  $r=N^{-1}\sum_{i}\sigma_{i}\sigma_{i+1}\sigma_{i}^\prime\sigma_{i+1}^\prime$.
Different rows correspond to different control parameters.
  Top row: high Poissonnian connectivity, viz. $J_{0}=0.25$ and $c=4$,  where the
 predicted equilibrium values are $m\simeq 0.75$, $q\simeq 0.58$, $a_1\simeq 0.62$, $a_2\simeq 0.57$, $r\simeq
 0.40$.
  Bottom row: low Poissonnian connectivity, viz. $J_{0}=1$ and $c=0.5$,  where the
theory  predicts $m\simeq 0.88$, $q\simeq 0.80$, $a\simeq 0.85$, $
a2\simeq 0.81$, $r\simeq 0.74$. In all cases the predictions are
indicated by markers at the right of the graphs. }
\label{fig:sim_sw}
\end{figure}

\begin{figure}[t]
\setlength{\unitlength}{0.6mm}\hspace*{12mm}
\begin{picture}(200,170)
\put(0,80){\includegraphics[height=77\unitlength,width=80\unitlength]{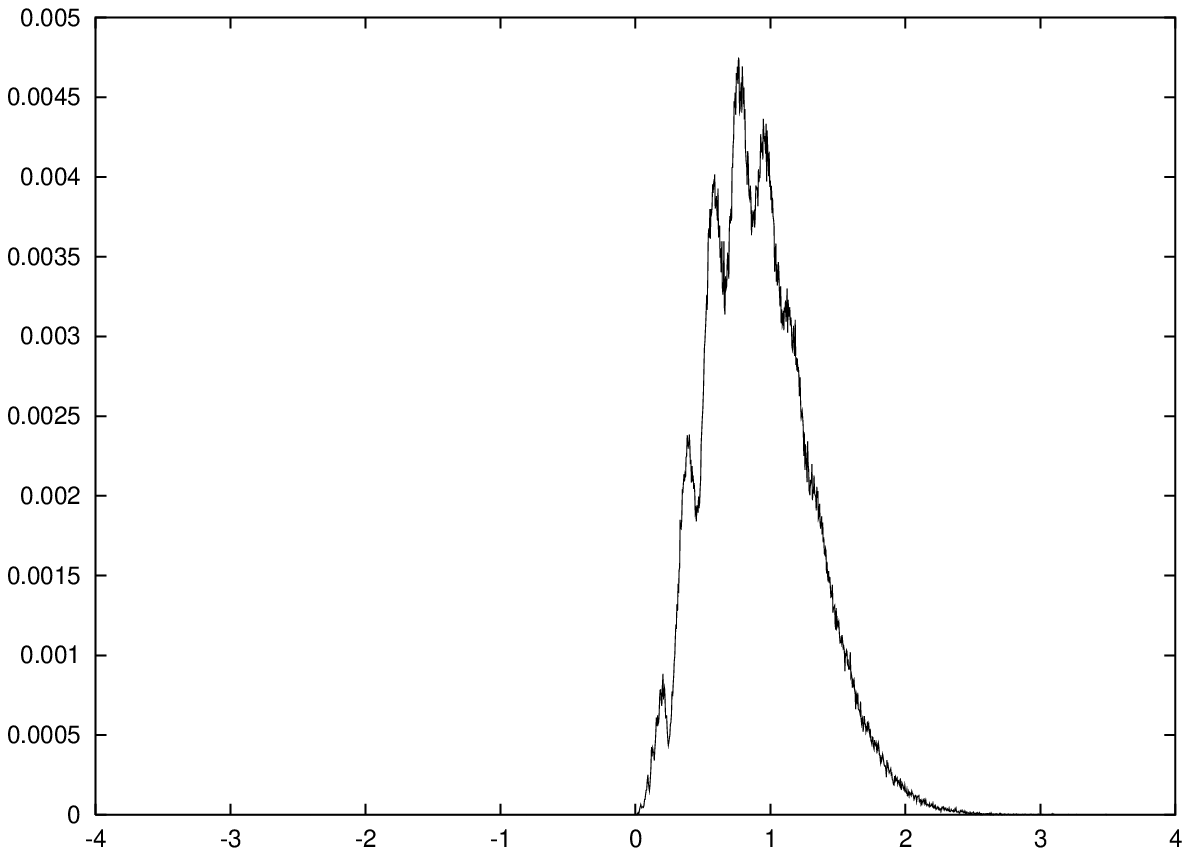}}
\put(80,80){\includegraphics[height=77\unitlength,width=80\unitlength]{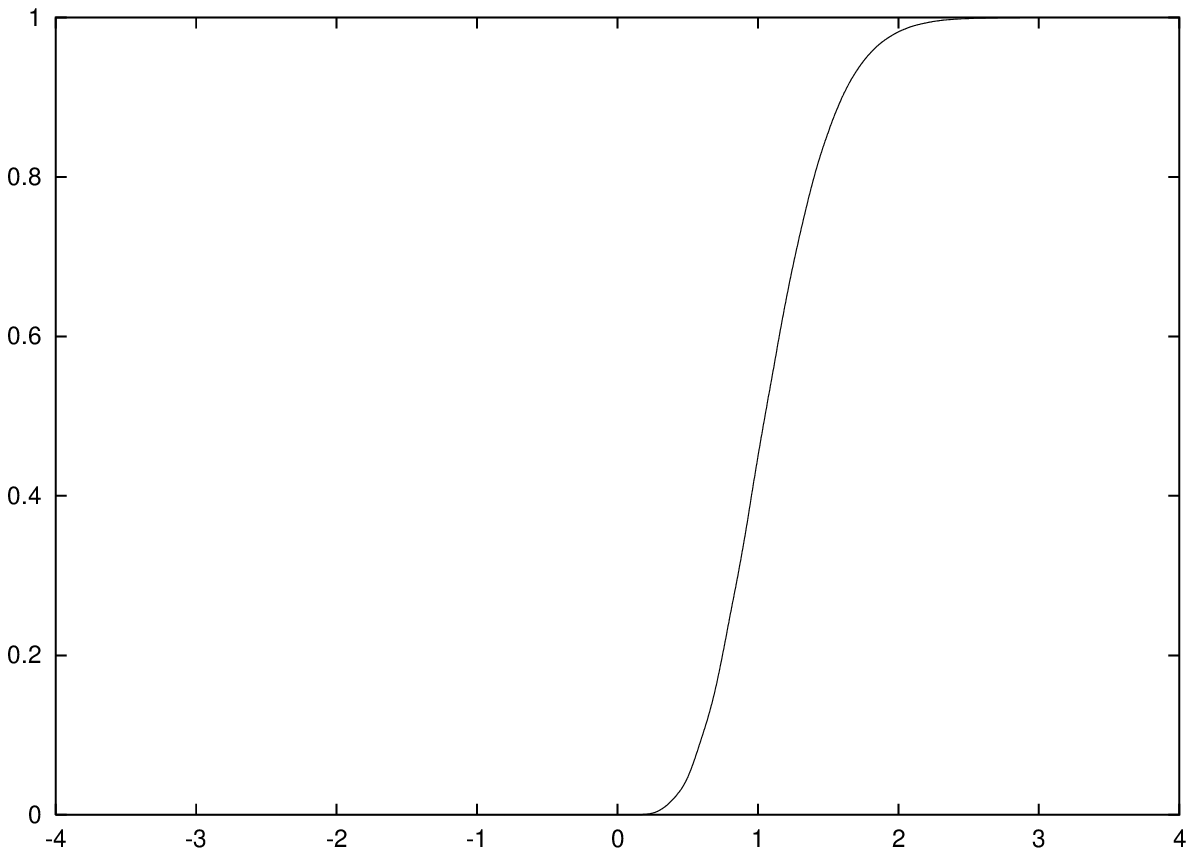}}
\put(160,80){\includegraphics[height=77\unitlength,width=80\unitlength]{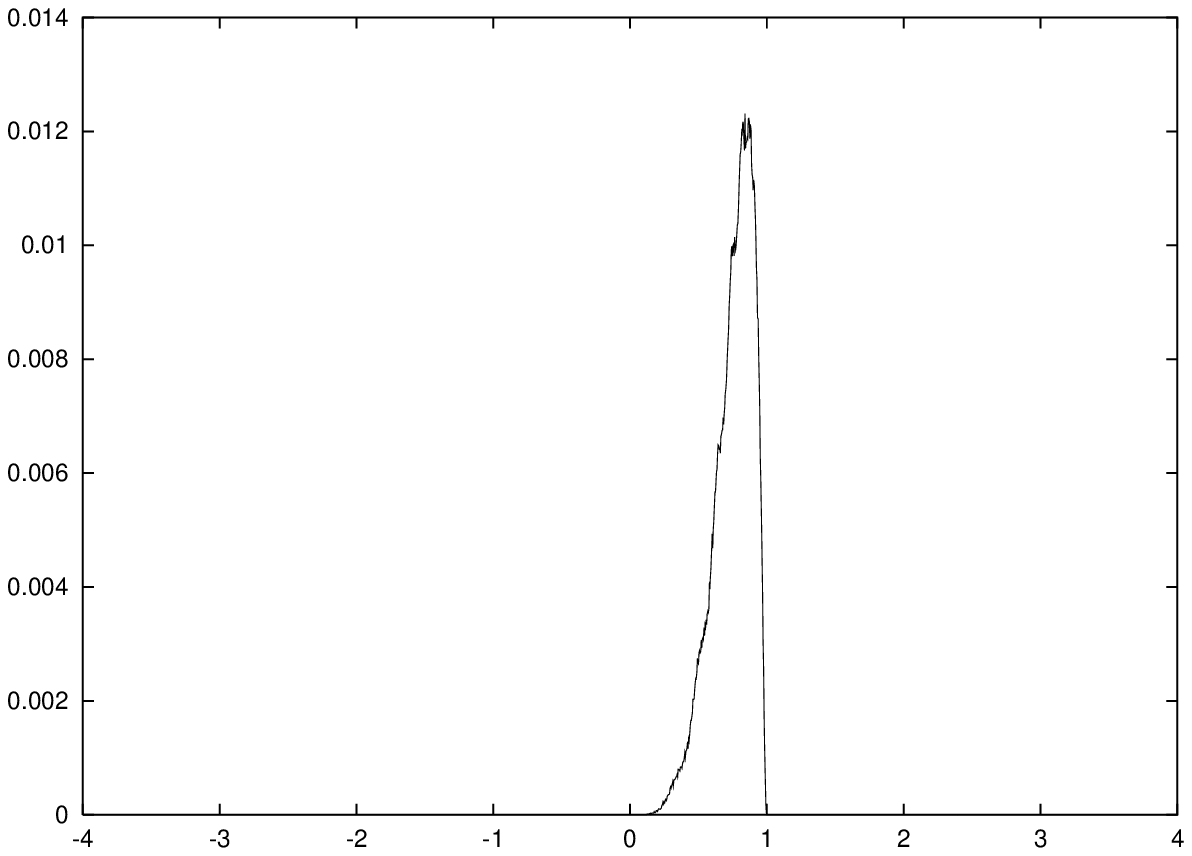}}
\put(0,0){\includegraphics[height=77\unitlength,width=80\unitlength]{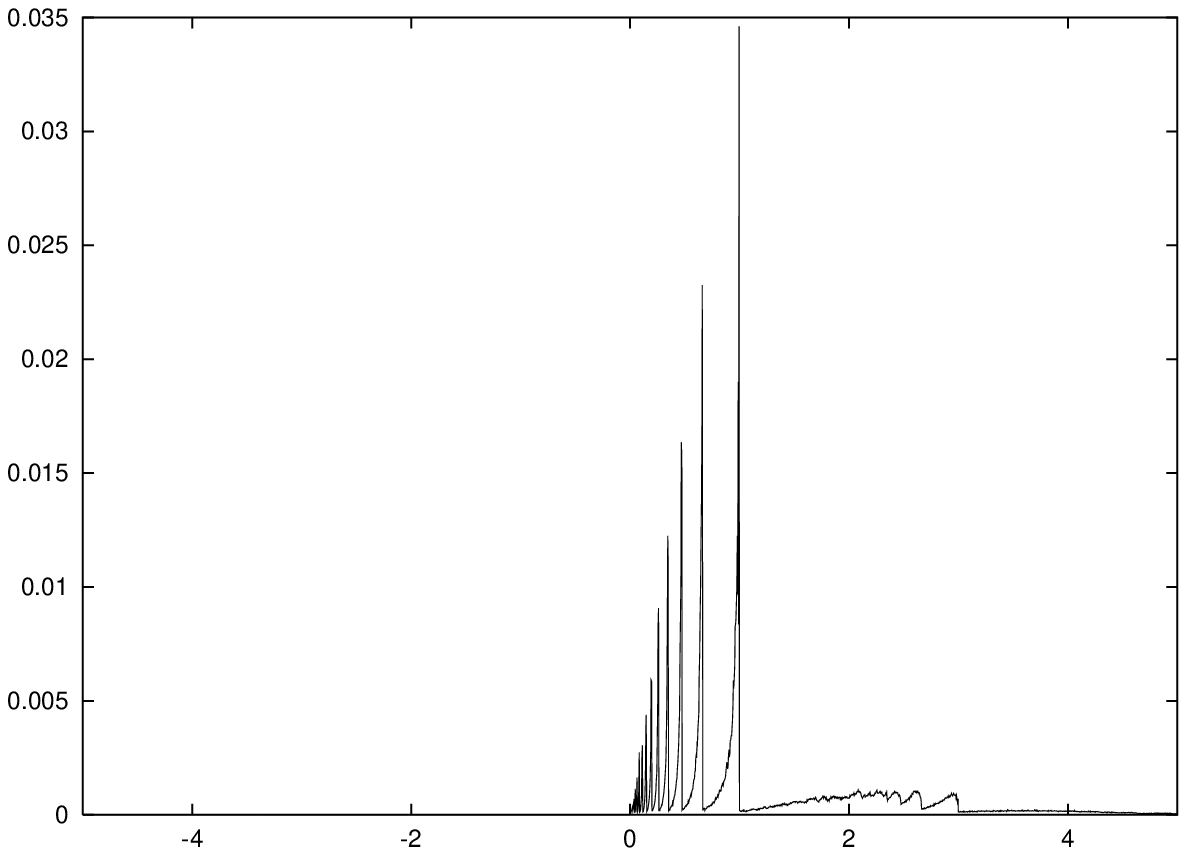}}
\put(80,0){\includegraphics[height=77\unitlength,width=80\unitlength]{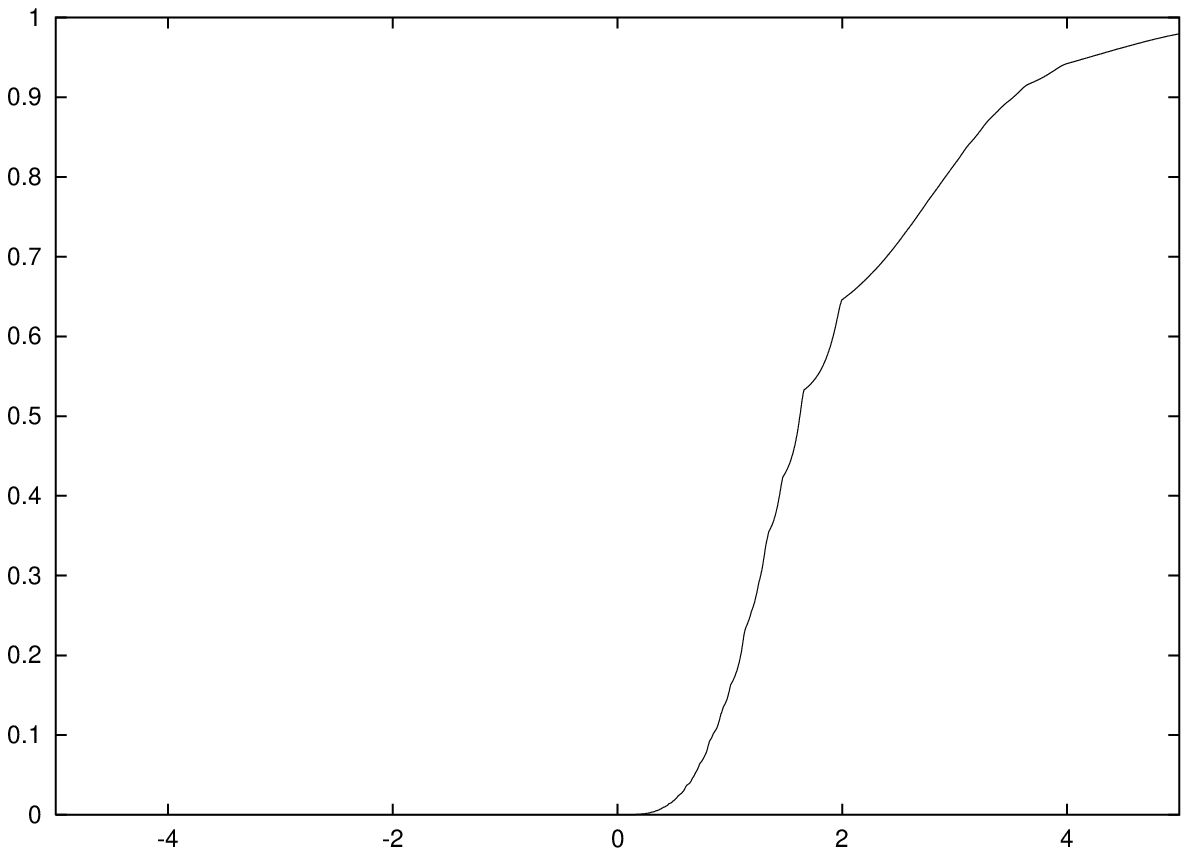}}
\put(160,0){\includegraphics[height=77\unitlength,width=80\unitlength]{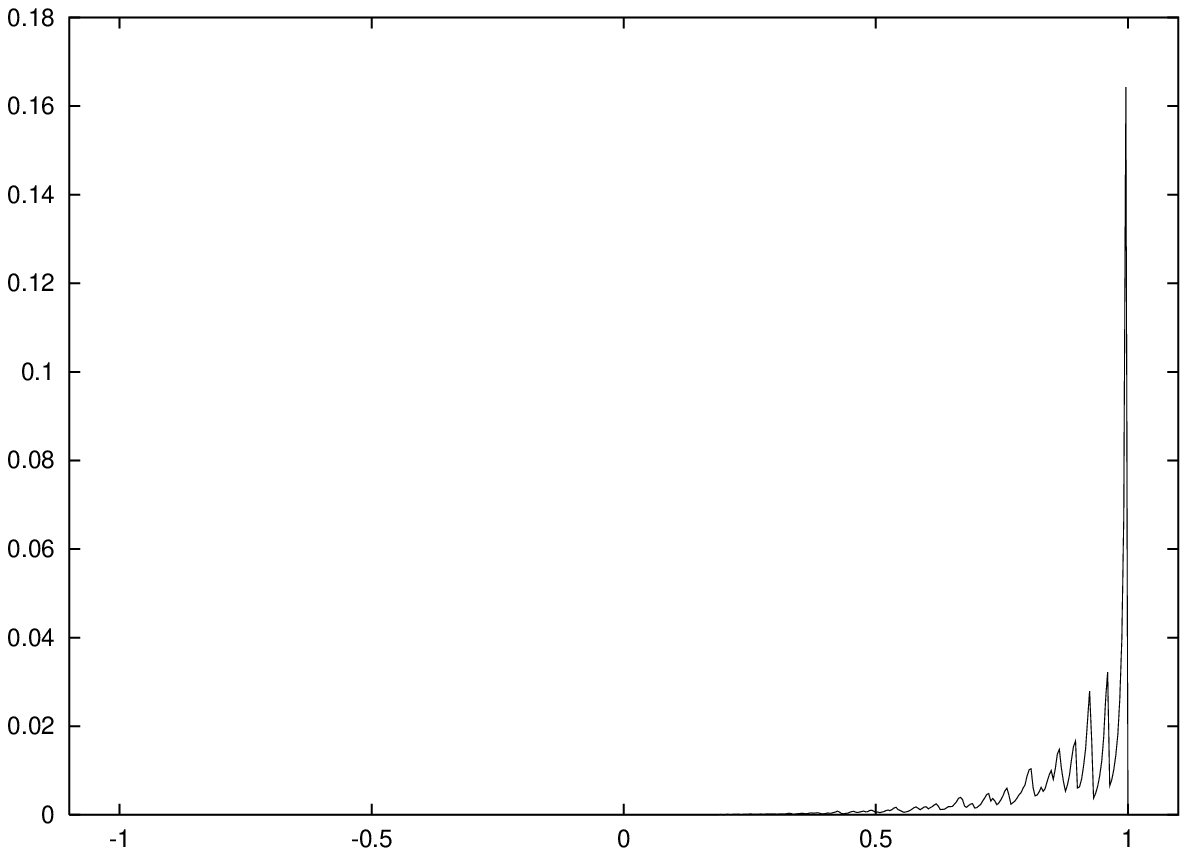}}
\put(42,-8){\here{\small $x$}} \put(122,-8){\here{\small$x$}}
\put(202,-8){\here{\small $m$}} \put(42,165){\here{$\Phi(x)$}}
\put(122,165){\here{$\hat{\Phi}(x)$}} \put(202,165){\here{$W(m)$}}
\end{picture}
\vspace*{8mm} \caption{ Field distributions corresponding to the
data of the previous figure \ref{fig:sim_sw}, obtained by
numerical solution of our integral eigenvalue equation (\ref{eq:
PHI}) via a population dynamics algorithm. The rows correspond to
examples with high  (top row) and low (bottom row) Possonnian
connectivity. Left column: The effective field distribution
$\Phi(x)$. Middle:
         the integrated distribution
         $\hat{\Phi}(x)=\int_{-\infty}^{x}\!\mathrm{d}z~\Phi(z)$.
         Right: the distribution of single-site magnetizations
         $W(m)=\int\!\mathrm{d}x\mathrm{d}y~\Phi(x)\Psi(y)\delta[m-\tanh(\beta x+\beta
         y)]$.}
\label{fig:th_sw1}
\end{figure}

Our final application example is the so-called 'small-world'
ferromagnet,  defined by the Hamiltonian (\ref{eq:SWH}). As in the
previous example  this model represents a combination of
one-dimensional short-range interactions and long-range ones. In
contrast to the previous example the long-range bonds are not
`all-to-all', but represent a finitely connected Poissonnian
random graph.  This model was studied in more detail in
\cite{nik-coolen04}, where it was shown that application of the
replica formalism generates the following replicated transfer
matrix, with $\bsigma,\bsigma^\prime,\btau\in\{-1,1\}^n$:
\be
\label{eq: rtm_sw}
  T(\bsigma,\bsigma^\prime\vert P)=
    e^{\beta J_{0}\bsigma\cdot\bsigma^\prime+
       c\sum_{\btau}P(\btau)\exp\Big[\frac{\beta J}{c}\bsigma\cdot \btau\Big]-c}
\ee Here the mean-field order parameter is a function $P(\btau)$,
which gives the fraction of sites where the replicated spin
$\bsigma_i$ equals $\btau$. The saddle-point equations are here
found to take the form of an expression for $P(\btau)$ in terms of
those eigenvectors of $\bT$ which correspond to the largest
eigenvalue:
\be
\label{eq: Psigma}
  P(\btau)=\frac{v_{0}(\btau)u_{0}(\btau)}
                {\sum_{\btau}v_{0}(\btau)u_{0}(\btau)}
\ee (assuming this eigenspace to be non-degenerated, similar to
our previous models). In this model one expects a replica
symmetric solution (RS) to describe the physics correctly, which
for the order parameter $P(\btau)$ implies the form
\be
  P(\btau)=\int\!\mathrm{d}h~W(h)\frac{e^{\beta h\sum_{\alpha=1}^{n}\tau_{\alpha}}}
                                     {[2\cosh(\beta h)]^{n}}
                                     \label{eq:RS}
\ee Insertion of this RS expression into (\ref{eq: rtm_sw})
results in the following replicated transfer matrix:
\be
\label{eq: rtm_sw_rs}
  T^{RS}(\bsigma,\bsigma^\prime)=
   \int\!\mathrm{d}\theta~p(\theta\vert n)~e^{\beta J_{0}\bsigma\cdot\bsigma^\prime+
           \beta\theta\sum_{\alpha}\sigma_{\alpha}}
\ee
\be
  p(\theta\vert n)=\sum_{k}\frac{e^{-c}c^{k}}{k!}\int
    \Big\{\prod_{r=1}^{k}\frac{\mathrm{d}h_{r}~W(h_{r})e^{n\beta B(J/c,h_{r})}}
                             {[2\cosh(\beta h_{r})]^{n}}\Big\}
   \delta[\theta-\sum_{r}A(J/c,h_{r})]
\ee
 Again we observe that our replicated transfer matrix
may be viewed as equivalent to that of a one-dimensional chain
with suitably chosen random fields. The associated `distribution'
of these fields represents the overall effect within the system of
the sparse Poissonian long range bonds on a given site of the
ring. We note that $p(\theta\vert n)$ is normalized only for
$n=0$.

Having identified the structure of our RS replicated transfer
matrix, one may proceed to solve this model using the eigenvectors
introduced in section \ref{sec:eigenvectors}. As in the Ising
chain, this results in a  transformation of the eigenvalue problem
to integral equations, viz. (\ref{eq:eigP},\ref{eq:kerP}) and
(\ref{eq:eigQ},\ref{eq:kerQ}),  involving now the above field
distribution $p(\theta\vert n)$. In addition the integral
eigenvalue  equations become coupled with the new distribution
$W(h)$ in (\ref{eq:RS}), which may be viewed as the fundamental
`mean-field' order parameter in this model.  In the limit $n\to 0$
one finds that $W(h)$ is given by
\be
  W(h)=\int\!\mathrm{d}x\mathrm{d}y~\Phi(x)\Psi(y)\delta(h-x-y)
\ee
 In order to find also correlation functions in the present
model we return to the previous derivation in section
\ref{sec:rfim} and invoke the identity: \bd
  \overline{\bra\sigma_{i}\sigma_{j}\ket^{\rho}}=
  \lim_{n\to 0}\sum_{\{\bsigma\}}\sigma_{1}^{\alpha_{1}}\sigma_{j}^{\alpha_{1}}\ldots
    \sigma_{i}^{\alpha_{\rho}}\sigma_{j}^{\alpha_{\rho}}
    \overline{\prod_{\alpha=1}^{n}e^{-\beta H(\bsigma^{\alpha})}}
\ed
 We find, after some straightforward and by now standard manipulations (viz. averaging over the
disorder, insertion of the relevant order parameters, and use of
saddle point equations) that correlation functions can be again
written in the form \bd
  \overline{\bra\sigma_{i}\sigma_{j}\ket^{\rho}}=
  \lim_{n\to 0}\frac{\tr(\bS_{\{\rho\}}\bT^{j-i}[P]\bS_{\{\rho\}}\bT^{N-j+i}[P])}{\tr(\bT^{N}[P])}
\ed where $P$ is now given by expression  (\ref{eq:RS}). Since the
steps which led us earlier for the random field Ising chain to
(\ref{eq:corlapp}) apply again,  we may simply use
(\ref{eq:corlapp}) again to find also the correlation functions
for the present model. The results of solving the relevant order
parameter equations numerically (via population dynamics
algorithms) are shown in  figure \ref{fig:sim_sw}, where we show
the predicted equilibrium values for the scalar observables
(\ref{eq:single_sys},\ref{eq:double_sys})
 together with the corresponding measurements
in numerical simulations, for comparison. The corresponding
effective field distributions are shown in figure
\ref{fig:th_sw1}.  As with the random field Ising model, the order
parameter functions required for the calculation of $m$ and $q$
have been calculated using the exact equations, whereas those
required for the multiple-site observables $\{a_1,a_2,r\}$ have
been solved approximately. This is borne out by figure
\ref{fig:sim_sw}, which indeed shows excellent agreement between
theory and simulations for $m$ and $q$ (left column), but
deviations for the three  quantities that have been calculated in
approximation (right column).


\section{Discussion}

In this paper we have developed new tools for the diagonalization
 of replicated transfer matrices, which arise upon applying the replica method
 to disordered models with one-dimensional short-range bonds,
 possibly in combination with (random) long range ones.
 Our method was based on mapping the problem of diagonalizing
 $2^n\times 2^n$ matrices which are invariant under the replica
 permutation group onto the problem of diagonalizing appropriate
 $n$-dependent integral operators, in which the limit $n\to 0$
 can be much more easily taken, via a suitable ansatz for the eigenvectors.
 The result, similar to that obtained earlier via more traditional methods,
 is an integral eigenvalue problem, which is exact in
 the relevant limits $N\to \infty$ and $n\to 0$, but which has to
 be solved numerically (using e.g. population dynamics).
 Given our explicit expressions for the eigenvectors, the route is open to the evaluation of the free
 energy and several families of disorder-averaged observables, including the
 magnetization and the spin-glass order parameter, but also
 multiple-site correlation functions. It should be emphasized, however, that to evaluate the
 latter types of objects we had to make two simplifying assumptions, for which the only basis as yet is their
 validity in simpler and thereby verifiable cases.

 We have developed our theory in full detail for the random field Ising
 chain, and we showed subsequently how the solution of  other more
 complicated models  can be obtained from this, especially those
 where short-range bonds are combined with
 long-range ones and where one effectively ends up with a random field Ising problem embedded within a
  mean-field calculation. In particular we have worked out our
 equations and predictions for $1+\infty$-dimensional recurrent neural networks,
 and for `small world' ferromagnets.

 Possible future applications of the approach
presented in this paper would be to the analysis of
two-dimensional disordered spin systems, or to models which
require finite$-n$ replica calculations (e.g. those where the
disorder is not truly frozen, but slowly and stochastically
evolving in time), or to situations where one has RSB (broken
replica symmetry) in $1+\infty$ dimensional or `small-world' spin
systems. The latter two  calculations would not seem to be easily
carried out using the more conventional random field methods as in
e.g. \cite{BrandtGross78,BruinsmaAeppli83}, but would appear to be
 feasible extensions of the procedures presented here.

\section*{Acknowledgment}

One of the authors (TN) acknowledges financial support from the State Scholarships
Foundation (Greece)

\appendix
\section{Combinatorial terms in the $n\to 0$ limit}

Here we prove identity (\ref{eq:combinatorics}). We note that the
natural continuation of factorials to non-integer values is via
the Gamma function \cite{Menzel}, viz. $n!=\Gamma(n+1)$. For
integer $\rho\geq 1$, integer $\ell> 0$  and real-valued $n<1$ (so
that always $\ell>n-\rho+1$) we may therefore write
\begin{eqnarray}
{n-\rho \choose \ell}&=& \frac{1}{\ell!}\lim_{\epsilon\downarrow
0} \frac{\int_\epsilon^\infty\!dx~x^{n-\rho}e^{-x}}
{\int_\epsilon^\infty\!dx~x^{n-\rho-\ell}e^{-x}}\nonumber
\\
&=&
 \frac{1}{\ell!}\lim_{\epsilon\downarrow 0}
\frac{\int_\epsilon^1\!dx~x^{n-\rho}e^{-x}+\order(\epsilon^0)}
{\int_\epsilon^1\!dx~x^{n-\rho-\ell}e^{-x}+\order(\epsilon^0)}
\nonumber
\\
&=&
\frac{1}{\ell!}\frac{n-\rho-\ell+1}{n-\rho+1}\lim_{\epsilon\downarrow
0} \frac{\epsilon^{n-\rho+1}+\order(\epsilon^0)}
{\epsilon^{n-\rho-\ell+1}+\order(\epsilon^0)} \nonumber
 \\&=&
\frac{1}{\ell!}\frac{n-\rho-\ell+1}{n-\rho+1}\lim_{\epsilon\downarrow
0} \frac{\epsilon^{\ell}+\order(\epsilon^{\ell+\rho-n-1})}
{1+\order(\epsilon^{\ell+\rho-n-1})} =0
\end{eqnarray}
We are left only with the case $\ell=0$, for which the above
factorial terms would be equal to one. This proves
(\ref{eq:combinatorics}).

\end{document}